\documentclass[twocolumn,reprint,amsmath,amssymb,footinbib,superscriptaddress,floatfix,aps,prl]{revtex4-1}
\pdfoutput=1
\usepackage{hyperref}
\usepackage[utf8]{inputenc}
\usepackage[sort&compress]{natbib}
\usepackage[dvipsnames]{xcolor}
\usepackage{graphicx}
\usepackage{epsfig}
\usepackage{epstopdf}
\usepackage{amsfonts}
\usepackage{amstext,amsmath,amssymb}

\usepackage{rotating}
\usepackage{dcolumn}
\usepackage{array,multirow}
\usepackage[english]{babel}
\usepackage[export]{adjustbox}
\usepackage{braket}
\usepackage[caption=false]{subfig}
\usepackage{xr}
\usepackage{ifpdf}
\usepackage{tikz}
\usetikzlibrary{arrows}

\hypersetup{urlcolor=violet, citecolor=violet, linkcolor=violet, colorlinks=true}
\newcommand{\be}{\begin{equation}}
\newcommand{\ee}{\end{equation}}
\newcommand{\bea}{\begin{eqnarray}}
\newcommand{\eea}{\end{eqnarray}}

\newcommand{\w}{\omega}
\newcommand{\W}{\Omega}

\newcommand{\prj}[1]{\left\vert #1    \right\rangle \left\langle   #1  \right\vert}

\usepackage{color}

\DeclareMathOperator{\sinc}{sinc}

\graphicspath{{figures/}}

\begin{document}

\title{Low field nano-NMR via three-level system control}

\author{J. Cerrillo}
\email{javier.cerrillo@upct.es}
\affiliation{\'Area de F\'isica Aplicada, Universidad Polit\'ecnica de Cartagena, Cartagena E-30202, Spain}
\author{S. Oviedo Casado}
\affiliation{Racah Institute of Physics, The Hebrew University of Jerusalem, Jerusalem 91904, Givat Ram, Israel}
\author{J. Prior}
\affiliation{Departamento de Física - CIOyN, Universidad de Murcia, Murcia E-30071, Spain}
\affiliation{\'Area de F\'isica Aplicada, Universidad Polit\'ecnica de Cartagena, Cartagena E-30202, Spain}
\affiliation{Instituto Carlos I de F\'isica te\'orica y Computacional, Universidad de Granada, Granada E-18071, Spain}

\date{\today}
\begin{abstract}
Conventional control strategies for NV centers in quantum sensing are based on a two-level model of their triplet ground state. However, this approach fails in regimes of weak bias magnetic fields or strong microwave pulses, as we demonstrate. To overcome this limitation, we propose a novel control sequence that exploits all three levels by addressing a hidden Raman configuration with microwave pulses tuned to the zero-field transition. We report excellent performance in typical dynamical decoupling sequences, opening up the possibility for nano-NMR operation in low-field environments.
\end{abstract}


\maketitle

Paramagnetic defects in diamonds feature prominently in magnetometry oriented quantum sensing. The most widely used of these defects is the negatively charged electronic spin of a nitrogen-vacancy center (NV)  \cite{Jelezko2002,Jelezko2004,Doherty2013}, whose ground state can be described in terms of a spin-1 triplet endowed with a surprisingly long coherence time even at ambient conditions \cite{Childress2006,Balasubramanian2008,Maze2008,Jacques2009}.
Its unique properties make of the NV center a remarkable sensor of the minute magnetic fields created by nano-sized samples \cite{Staudacher2013,Mamin2013,Muller2014,Ajoi2015,Lovchinsky2016,Glenn2018}. The NV has positioned itself as the leader in the race to achieve the ultimate nuclear magnetic resonance (NMR) spectrometer, in which a spin probe measures the magnetic field created by a distant single spin \cite{Lukin2012,Hanson2012,Zhao2012,Laraoui2013}. As well, a superior ability to measure other physical magnitudes such as temperature or electric fields has been demonstrated \cite{Dolde2011,Neumann2013}. The success of NV centers as quantum sensors is based on the ability to control the spin probe and extend the duration of its coherence beyond the correlation time of the measured oscillating signal.

\begin{figure}
\centerline{
    \begin{tikzpicture}[
      level/.style={very thick},
      axes/.style={thin},
      virtual/.style={ultra thin,densely dashed},
      transthb/.style={ultra thick,blue!60!,<->,shorten >=2pt,shorten <=2pt,>=stealth},
      transthbl/.style={ultra thick,black,<->,shorten >=2pt,shorten <=2pt,>=stealth},
      transb/.style={ultra thin,blue!50!black,<->,shorten >=2pt,shorten <=2pt,>=stealth},
      transbr/.style={ultra thin,red!20!,<->,shorten >=2pt,shorten <=2pt,>=stealth},
      transbl/.style={ultra thin,densely dashed,<->,shorten >=2pt,shorten <=2pt,>=stealth},
      transthr/.style={ultra thick,color=red!80!black,<->,shorten >=2pt,shorten <=2pt,>=stealth},
    ]
    \node[] (russell) at (-1cm,9em) {a)};
    \draw[axes] (0cm,-2em) -- (4cm,-2em) node[right] {$B$};
    \draw[axes] (0cm,-2em) -- (0cm,9em) node[left] {$E$};
    \draw[level] (0cm,0em) -- (4cm,0em) node[right] {$\ket{0}$};
    \draw[level] (0cm,6.5em) -- (4cm,5em) node[right] {$\ket{-1}$};
    \draw[level] (0cm,6.5em) -- (4cm,8em) node[right] {$\ket{+1}$};
    \draw[virtual] (1.4cm,6.5em) -- (-.3cm,6.5em);
    \draw[virtual] (0cm,0em) -- (-.3cm,0em);
    \draw[transthb] (1.2cm,-0.2em) -- (1.2cm,6.7em)  node[midway,left] {$\nu=D$};
    \draw[transthr] (3.5cm,-0.2em) -- (3.5cm,5.4em) node[midway,left] {$\nu=D-\mu B$} ;
    \draw[transbl] (3.5cm,5em) -- (3.5cm,8em) node[midway,right] {$2 \mu B$};
    \draw[transbl] (-.2cm,0em) -- (-.2cm,6.5em) node[midway,left] {$D$};
    \node[] (russell) at (-2.3cm,-5em) {b)};
    \draw[level] (-.8cm,-10em) -- (-.3cm,-10em) node[right] {$\ket{0}$};
    \draw[level] (-2cm,-6em) -- (-1.5cm,-6em) node[right] {$\ket{-1}$};
    \draw[level] (0cm,-5em) -- (.5cm,-5em) node[right] {$\ket{+1}$};
    \draw[transthr] (-.6cm,-10em) -- (-1.8cm,-6em);
    \draw[transthr] (-.6cm,-10em) -- (.3cm,-6em);
    \draw[virtual] (0cm,-6em) -- (.5cm,-6em);
    \node[] (russell) at (4cm,-8em)
    {\includegraphics[width=.5\columnwidth]{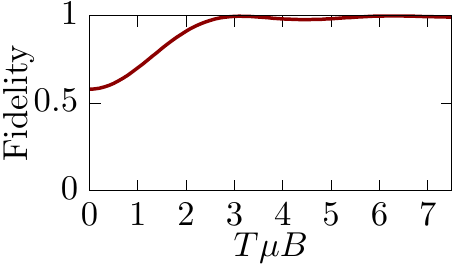}};
    \node[] (russell) at (-2.3cm,-13em) {c)};
    \draw[level] (-.5cm,-19em) -- (0cm,-19em) node[right] {$\ket{0}$};
    \draw[level] (-1cm,-15em) -- (-1.5cm,-15em) node[left] {$\ket{-1}$};
    \draw[level] (0.3cm,-13em) -- (.8cm,-13em) node[right] {$\ket{+1}$};
    \draw[transthb] (-.3cm,-19em) -- (-1.3cm,-14em);
    \draw[transthb] (-.3cm,-19em) -- (.6cm,-14em);
    \draw[virtual] (0.3cm,-14em) -- (.8cm,-14em);
    \draw[virtual] (-1.5cm,-14em) -- (-1cm,-14em);
    \node[] (russell) at (2cm,-17em) {\Large $\iff$};
    \draw[level] (4.2cm,-19em) -- (4.7cm,-19em) node[right] {$\ket{0}$};
    \draw[level] (4cm,-14em) -- (3.5cm,-14em) node[left] {$\ket{-}$};
    \draw[level] (4.8cm,-14em) -- (5.3cm,-14em) node[right] {$\ket{+}$};
    \draw[transthbl] (4cm,-14em) -- (4.8cm,-14em) node[midway, below] {$\mu B$};
    \draw[transthb] (4.4cm,-19em) -- (5cm,-14em) node[midway, right] {$\frac{\W}{2}$};
    \end{tikzpicture}
}
\caption{
(a) The ground state manifold of the NV is a spin-1 triplet where the $\ket{\pm 1}$ states are degenerate and separated from the state $\ket{0}$ by a zero field splitting of $D = 2.87$GHz. A static magnetic field produces further splitting of the $\ket{\pm 1}$ states and allows to work with a quasi 2LS, via DD sequences of frequency matching the $\ket{0} \rightarrow \ket{-1}$ gap (red arrow). We propose instead addressing the whole 3LS with a DD frequency $\nu = D$ (blue arrow). (b) Fidelity of state $\left(\ket{-1} + i\ket{0}\right)/\sqrt{2}$ after a microwave $\pi/2$ pulse of duration $T$ applied on an NV-center initialized in state $\ket{0}$, represented as a function of the dimensionless quantity $T\mu B$. At small Zeeman splittings or short pulse lengths, poor fidelities are achieved due to off-resonant excitation of the state $\ket{+1}$. (c) By driving the system at $\nu=D$, an effective Raman coupling is implemented, providing accurate control of the system even at large intensities or low bias fields.}
\label{levels}
\end{figure}
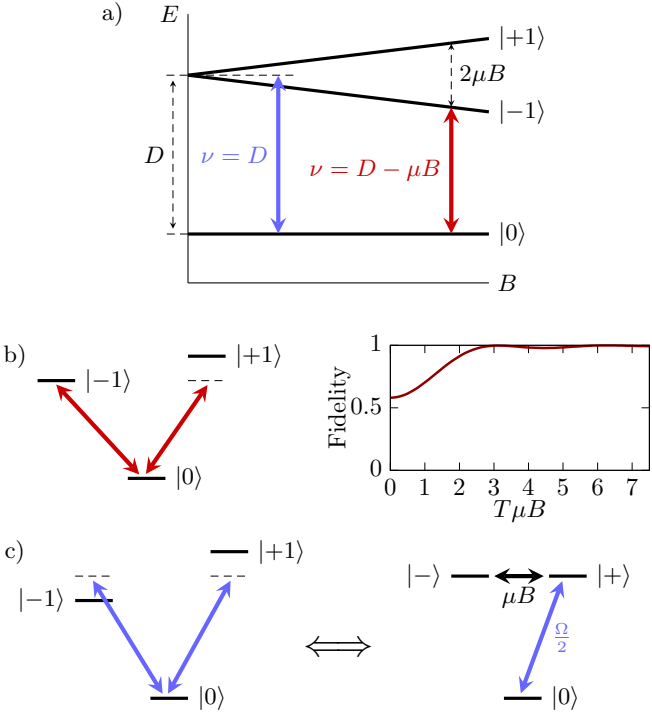

Typically, the NV center is considered as a two level system (2LS) thanks to the Zeeman splitting of the $\ket{\pm 1}$ levels that is introduced by means of a bias magnetic field aligned with the NV axis. Most control protocols involve dynamical decoupling (DD) sequences of microwave pulses resonant with the $\ket{0} \leftrightarrow \ket{-1}$ transition \cite{Maudsley1986,Souza2011,Wang2011,Kotler2011,Souza2012, Cywinski2007,Uhrig2008,Uhrig2007a,deLange2010,Ryan2010,Naydenov2011,Rotem2019,Casanova2015,Abobeih2018,Wang2019}, as indicated by the red arrow in Fig.\ref{levels}a.
These protocols conventionally make use of the assumption that pulses are negligibly short (impulsive limit). This actually poses a problem since, in order to conserve the pulse area, the intensity must compensate for the reduced duration. When the intensity becomes comparable to the Zeeman splitting, the 2LS approximation fails \cite{London2014}, resulting in a degradation of coherence time and sensitivity. Nevertheless, working in the regime of very low Zeeman splitting ($\lesssim 20 \mathrm{G}$) becomes necessary for ultra-low field detection or environment quantum control \cite{Reinhard2012,Laraoui2013,Walsworth2018,Schmitt2017,Ajoi2019,Zheng2019,Walsworth2020}.

In this Letter, we propose an alternative route to high precision low-field quantum sensing through three-level system (3LS) techniques  \cite{Cerrillo2010,Cerrillo2018}. Rather than struggling with the 2LS approximation
we utilize a different DD frequency which equally addresses all of the NV ground spin triplet states, as depicted in Fig.\ref{levels}. This allows high precision measurements at low Zeeman splitting or, equivalently, permits the use of high-power pulses, consequently providing a much wider sensitivity range. Previous works involving the $\ket{+1}$ spin state have already demonstrated  
fourfold enhanced sensitivity thanks to the double quantum transition $\ket{-1}\leftrightarrow\ket{+1}$
\cite{Watanabe2013,Mamin2014,Walsworth2018,Casanova2020}. However, these proposals still rely, just as conventional methods, on the 2LS approximation, which brings back the problem of state overlap at low bias field. Instead, by addressing both $\ket{\pm 1}$ states with the same pulse, our proposal manages accurate 3LS control with stark robustness to pulse errors, thus providing high fidelity in a regime so far lacking appropriate control protocols.

Moreover, we tackle the effect of the finite length $T$ of the pulses. This has previously been considered in the context of DD sequences, demonstrating that the finite temporal width of the pulses is crucial to calculate the optimal phase acquisition time and to match the spin precession with the signal frequency \cite{Bollinger2009,Ishikawa2018}. Nonetheless, these approaches focus on 2LSs and assume complete suppression of the signal whenever the pulse is acting, ignoring that some phase is indeed gathered during that time. Instead, we provide an improved analytical approach and full numerical calculations.

The combination of a Raman control strategy \cite{Cerrillo2010} for NV centers together with the consideration of finite duration of pulses opens up the possibility of highly precise weak measurements in low field nano-scale NMR. 

\emph{Conventional control strategy} ---
We consider an NV center under the effect of an external, bias magnetic field in the $z$ direction $\mathbf{B}=(0,0,B)$ and controlled with a microwave field of frequency $\nu$. The corresponding Hamiltonian is
\be
H(\nu)= D S_z^2 + \mu B S_z + \W \cos \nu t S_x,
\label{Hw}
\ee
where $D$ is the zero-field splitting, $\mu$ is the gyromagnetic ratio of electrons, $\W$ is the Rabi frequency associated with the microwave drive and $S_j$ is the spin-1 Pauli matrix in the $j$th direction. The dependence of the eigenvalues with the strength of the magnetic field $B$ is depicted in Fig.\ref{levels}a.

Control of the NV center is usually achieved by setting $\nu=D-\mu B$ (red coupling in Fig.\ref{levels}a). We will refer to this strategy as the conventional control (CC) scheme. At this frequency, the microwave field spectrally addresses the $\ket{0}\leftrightarrow\ket{-1}$ transition, as revealed by the Hamiltonian in the appropriate rotating picture
\bea
\tilde{H}(D-\mu B) &\simeq &2 \mu B \prj{+1}\\
 &&+ \frac{\W}{2\sqrt{2}}\left(\ket{-1}\bra{0}+\ket{+1}\bra{0}+H.c.\right).\nonumber
\eea
Due to the existence of a matrix element $\bra{0}S_x\ket{+1}$, the 2LS approximation remains valid only as long as transitions to the state $\ket{+1}$ can be neglected, either because the bias field $B$ sets it sufficiently apart or because the microwave intensity $\W$ is weak enough. Beyond this regime, pulse control becomes imprecise and fails to perform the expected task, as illustrated in Fig.\ref{levels}b for the case of a $\pi/2$-pulse of duration $T=\pi/(\sqrt{2}\W)$. There, we represent the fidelity of the target state $\left(\ket{-1} + i\ket{0}\right)/\sqrt{2}$ as a function of the dimensionless product $T\mu B$. The target state becomes depleted by off-resonant excitation to $\ket{+1}$ either as the Zeeman splitting is reduced or as the pulse intensity increases. Therefore, the ideal of the impulsive limit is actually not desirable in this context and it is always necessary to resort to finite pulse lengths.

\begin{figure*}
    \includegraphics[trim=0 9.3cm 0 0, clip,width=1\linewidth]{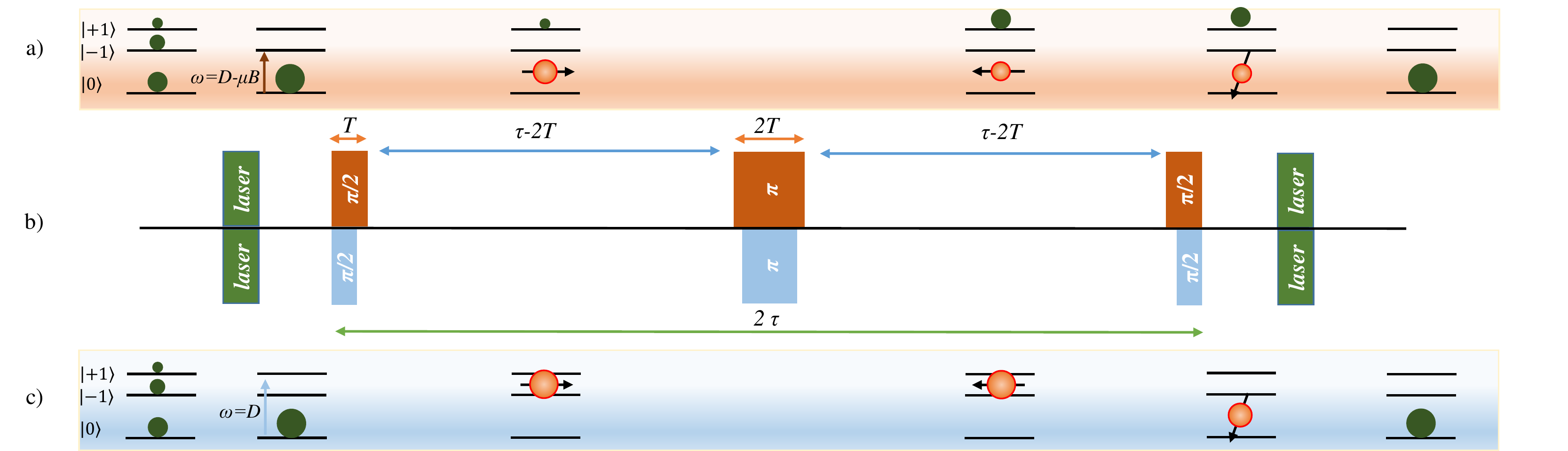}
    \caption{\emph{Protocol} - Depiction of a DD protocol for magnetometry with NV centres and comparison between CC and NV-ERC. In a), the usual DD sequence (upper b) creates an initial superposition $\ket{0} + \ket{-1}$, but additionally partially populates the $\ket{+1}$ state. Pulses have to be lengthened in order to ensure a high fidelity truer to the 2LS framework. As a consequence, sensitivity is limited. In c) on the contrary, the pulses address the three levels and create a coherent superposition $\ket{+1} + \ket{-1}$ without contamination from the $\ket{0}$ state. This permits reducing the duration of the pulses in our proposed DD sequence (lower b), which increases sensitivity for the same total duration, or permits introducing more pulses within the same DD sequence, without loss of signal and working at low bias field.}
    \label{pulses}
\end{figure*}

\emph{Proposed control strategy} ---
In order to avoid the detrimental effect illustrated in Fig.\ref{levels}b, we propose the effective Raman control (NV-ERC) strategy, which involves both the $\ket{-1}$ and $\ket{+1}$ states by setting the microwave frequency rather to $\nu=D$ (blue coupling in Fig.\ref{levels}a). In the absence of noise sources, a coherent superposition with fidelity 1 is possible at this spectral position even in the limit of high intensity pulses or small Zeeman splitting (for the effect of MW source noise see supplementary material \cite{SM}). This is due to the hidden effective Raman coupling
\be
\tilde{H}(D)\simeq  \left(\mu B \ket{-} + \frac{\W}{2}\ket{0} \right)\bra{+}+ H.c.,
\ee
revealed by the basis $\ket{\pm}=(\ket{+1}\pm\ket{-1})/\sqrt{2}$ (see Fig.\ref{levels}c). This Hamiltonian produces Rabi oscillations of frequency $\bar\W=\sqrt{\mu^{2}B^{2}+\Omega^{2}/4}$ and period $\bar{T}=2\pi /\bar\W$ between states $\ket{+}$ and $\left(\mu B \ket{-} + \frac{\W}{2}\ket{0} \right)/\bar\W$.

With this structure, it is possible to improve the two types of pulses required in most DD sequences. For the population to coherence map (commonly a $\pi/2$-pulse of length $T$), we introduce the notion of a pulse of strength $\W\ge 2\mu B$ applied for a time
\be
\bar{T}'=\frac{\arccos\left(-\frac{4 \mu^2 B^2}{\W^2}\right)}{\bar\W},
\ee
which transforms $\ket{0}$ into $\ket{\phi}=(\ket{-1} -\exp(i\phi)\ket{+1})/\sqrt{2}$ with $\cos\phi=\frac{8\mu^{2}B^{2}}{\W^{2}}-1$. The opposite transformation $\ket{\phi}\rightarrow\ket{0}$ is achieved in the time remaining off the Rabi period $\bar{T}''= \bar{T}-\bar{T}'$. Unlike the pulses discussed above, these remain exact for increasing $\W$.

The Bloch sphere rotation  (usually a $\pi$-pulse of length $2T$) is implemented by a $2\pi$-pulse with $\W\gg\mu B$ of length $T'=2\sqrt{2}T$, which produces the phase gate $\ket{+}\rightarrow-\ket{+}$. This pulse improves its performance as the impulsive limit is approached, also in stark contrast to the case described above.

\emph{Application in DD sequences} --- A schematic representation and comparison of both CC and NV-ERC strategies in the context of a Hahn-echo sequence is illustrated in Fig.\ref{pulses}. Both pulse sequences are represented in Fig.\ref{pulses}b. Additionally, the effects in the state of the 3LS are represented in Fig.\ref{pulses}a for CC and  Fig.\ref{pulses}c for NV-ERC. There, coherences between levels are represented by orange Bloch spheres and their corresponding vectors, while residual population is represented by green spheres of varying diameters. Starting with a thermal distribution of population across the three states, a laser pulse firstly initializes the NV at $\ket{0}$. A microwave $\pi/2$-pulse of length $T$ ($\bar{T}'$) then attempts to generate an equal superposition of the two states $\ket{0}$ and $\ket{-1}$ ($\ket{+1}$ and $\ket{-1}$). Unwanted population appears at $\ket{+1}$ in Fig.\ref{pulses}a, whereas this problem is avoided in Fig.\ref{pulses}c. After a free evolution time $\tau - 2T$ ($\tau-T'/2-\bar T/2$), where $2\tau$ is the total time of the sequence, a $\pi$-pulse of length $2T$ ($T'$) is applied to invert the Bloch sphere in the respective 2LS. Following an equal free evolution time, another $\pi/2$-pulse of duration $T$ ($\bar T''$) aims at converting coherences to populations, which are finally measured by a laser pulse (which also re-initializes the NV center). Each microwave pulse in Fig.\ref{pulses}a accumulates additional population in $\ket{+1}$, rendering the sequence increasingly imprecise.

{\em Ramsey sequence} --- To compare both strategies, we resort to the concept of the filter function (FF) $F(\w)=f_r^2(\w) + f_i^2(\w)$. For a coherent external field perturbation $\mathbf{B}_r'(\w)=\cos(\w t)(0,0,1)$ [$\mathbf{B}_i'(\w)=\sin(\w t)(0,0,1)$], $f_r(\w)$ [$f_i(\w)$] is equivalent to the accumulated phase in the NV \cite{SM}, allowing us to directly relate analytical calculations of FFs with numerical results of the performance of each control sequence. Let us first consider the simpler case of an ideal Ramsey experiment, which involves an initial $\pi/2$-pulse, waiting time and a final $\pi/2$-pulse. In the impulsive limit, it features the FF
\begin{equation}
 F_{R}(\omega) = \tau^2 \sinc^2\left(\frac{\omega \tau}{2}\right),
 \label{Rideal}
\end{equation}
where $\tau$ is the total duration of the experiment and $\sinc(x)=\sin(x)/x$. As expected, it peaks at vanishing frequency, confirming that this sequence is mostly sensitive to continuous noise. This function also indicates that longer experiments are preferred for two reasons. First, they yield stronger signals as indicated by the factor $\tau^2$. Second, longer experiments have better frequency selectivity as indicated by the $\sinc$-dependency. This is explained by the fact that only perturbations whose periods are longer than $\tau$ can be detected, while the signal associated to faster oscillations averages out.

As discussed above, the impulsive limit $T\rightarrow 0$ is not desirable in the context of NV centers. Nevertheless, a finite pulse length $T$ has detrimental effects in the sensitivity of experiments \cite{Bollinger2009,Ishikawa2018}.
This can be shown even in the context of an ideal 2LS, where an improved expression incorporating the dependency on $T$ can be derived \cite{SM}
\begin{equation}
 F_{R}(\omega,T) = \left(\tau-T\right)^2  \sinc^2\left(\w\frac{T}{2}\right)\sinc^2\left(\w\frac{\tau-T}{2}\right).
 \label{FR}
\end{equation}
The prefactor $\left(\tau-T\right)^2$ clearly indicates that the strength of the signal is heavily diminished by $T$. Additionally, a slight broadening of the function takes place between the ideal limit $T=0$ in Eq.\ref{Rideal} and the maximum pulse length possible $T=\tau/2$, where  $ F_{R}\left(\omega,\tau/2\right) = \left(\tau/2\right)^2  \sinc^4\left(\w\tau/4\right)$, so pulse length also reduces frequency selectivity.

\begin{figure}
    \includegraphics{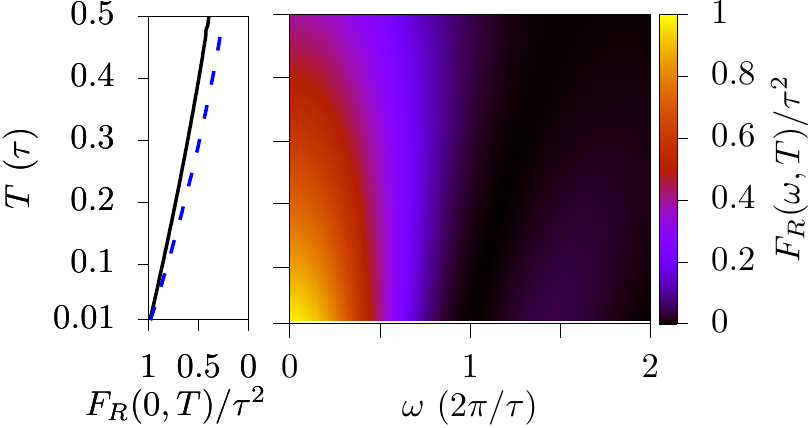}
    \caption{Filter function $F_R(\w,T)$ of a Ramsey experiment as a function of the $\pi/2$-pulse duration $T$ and the signal frequency $\w$ in the ideal case of a 2LS. The left panel compares the values of $F_R(0,T)$ from the numerical calculation (black solid curve) and the analytical calculation Eq.\ref{FR} (blue dashed curve).}
    \label{Ramsey2LS}
\end{figure}

Improving on Eq.\ref{FR}, we perform a numerical calculation of the remaining ground state population $p_{r,i}(\w,T)$ of a 2LS after a Ramsey sequence of total duration $\tau$ (including pulse length $T$) under the effect of the signal $\mathbf{B}_{r,i}'(\w)$. Using the relation $f_{r,i}=\arccos(1-2p_{r,i})$ between population and accumulated phase, we can compute the associated numerical FF as shown in Fig.\ref{Ramsey2LS}. The left panel represents the value of the FF for $\w=0$ as compared to the prediction of Eq.\ref{FR} $F_R(0,T)=(\tau-T)^2$. Signal loss with $T$ is slightly overestimated by Eq.\ref{FR}, but the trend is confirmed numerically as the most crucial consequence of lengthening the pulse duration. Frequency selectivity is also lost as shown by the slight broadening of the function with $T$, visible in the main panel.

\begin{figure}
    \includegraphics{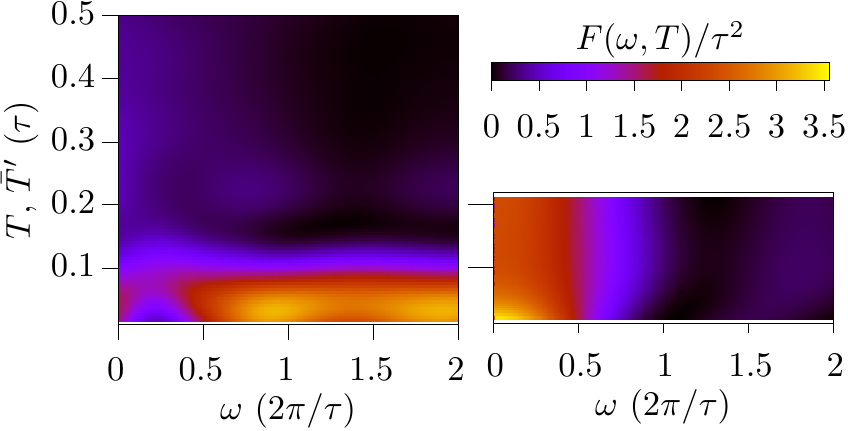}
    \caption{Filter function $F_R(\w,T)$ of a Ramsey experiment as a function pulse duration and signal frequency $\w$ for an NV center (Eq.\ref{Hw}) using CC (left panel, vertical axis corresponding to $T$) and NV-ERC (right panel, vertical axis corresponding to $\bar T'$). The value of the Zeeman splitting is $\mu B= 10/\tau$, which imposes $\bar{T}'\leq \tau \pi /10\sqrt{2}\simeq0.22\tau$.}
    \label{Ramsey3LS}
\end{figure}

An analogous numerical simulation has been performed for an NV center modeled by Eq.\ref{Hw} and the results are shown in Fig.\ref{Ramsey3LS} for CC (left panel) and for NV-ERC (right panel). Accumulated contamination of $\ket{+1}$ for $T\mu B \leq 1$ utterly blurs the signal in the case of CC. Only for $T\mu B \gg 1$ a faint Ramsey profile may be recovered. In contrast, NV-ERC strongly features the profile of Eq.\ref{Rideal} in the regime where standard operation fails. Additionally, it provides a rather intense signal due to the double quantum separation between the involved states $\ket{\pm 1}$.

{\em Hahn-echo sequence and beyond} --- More complex DD sequences are routinely used in experiment. They amount to the introduction of $N$ equally spaced $\pi$-pulses of length $2T$. The case $N=1$, known as the Hahn echo experiment, is illustrated in Fig.\ref{pulses}. Following the analysis above, an odd amount $N=2n+1$ of $\pi$ pulses produces a FF of the form \cite{SM}
\begin{equation}
 F_{2n+1}(\omega,T) =4\sin^2\left(\omega \frac{\tau}{2}\right)\frac{\cos^2\left(N\omega\frac{2\tau-T}{2}\right)}{\cos^2\left(\omega\frac{2\tau-T}{2}\right)} F_{R}(\omega,T),
 \label{Fodd}
\end{equation}
while an even amount $N=2n>0$ produces
\begin{equation}
 F_{2n}(\omega,T) =4\sin^2\left(\omega \frac{\tau}{2}\right)\frac{\sin^2\left(N\omega\frac{2\tau-T}{2}\right)}{\cos^2\left(\omega\frac{2\tau-T}{2}\right)} F_{R}(\omega,T).
\end{equation}
The first factor centers the signal around $\w=\pi/\tau$, whereas the second one centers it around  $\w=\pi/(\tau-T)$. The function becomes increasingly peaked around this last value for increasing $N$. Therefore, in addition to the effects discussed above, the pulse length $T$ disturbs the value of the addressed frequency. This behavior is reproduced by the numerical calculation for a Hahn-echo sequence on an ideal 2LS \cite{SM}. As in Fig.\ref{Ramsey3LS}, the same sequence on an NV center modeled by Eq.\ref{Hw} fails in the region $T\mu B \leq 1$, whereas with NV-ERC the Hahn-echo profile is successfully recovered in that regime. 

\begin{figure}[b]
    \includegraphics[width=\columnwidth]{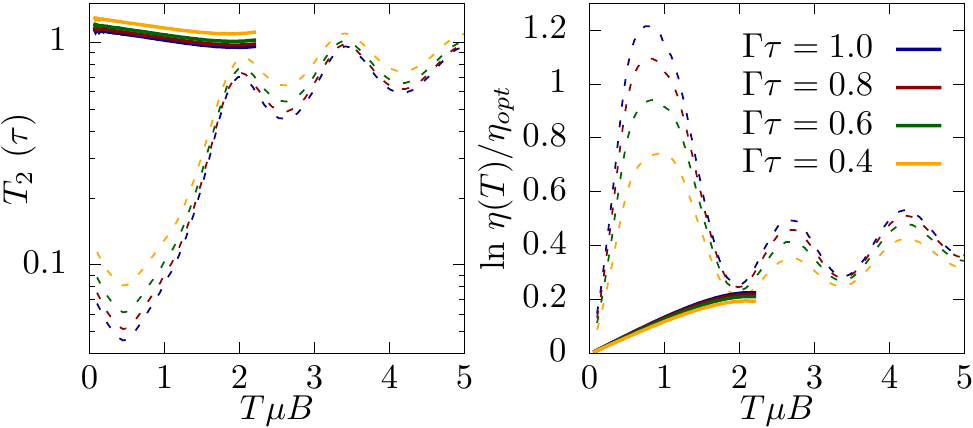}
    \caption{Estimated coherence time $T_2$ (left) and sensitivity $\eta$ as a function of the product $T\mu B$ for CC (dashed lines) and NV-ERC (solid lines) for different widths $\Gamma$ of a Lorentzian noise. Sensitivity is expressed in units of optimal sensitivity $\eta_{opt}$ in the impulsive limit and without decoherence effects.}
    \label{CohSens}
\end{figure}

In order to illustrate this effect in the context of experimentally relevant figures of merit, in Fig.\ref{CohSens} we estimate the corresponding coherence times $T_2$ and sensitivities $\eta$ \cite{SM} for both protocols in the presence of different Lorentzian environments. In the CC case, both coherence time and sensitivity are heavily degraded for small $T\mu B$, that is, in the presence of either small magnetic field or short, intense pulses. In contrast, the problem is resolved by NV-ERC in its region of applicability.

{\em Conclusions} --- Quantum sensing with arbitrary precision requires from an ever increasing sophistication of control protocols. Here we have demonstrated that the usual paradigm that considers the NV as a two-level system fails whenever the control pulses are strong or the bias field is low. We solve the problem by exploiting a Raman configuration involving both $\ket{\pm 1}$ states, and demonstrate that with such a control technique it is possible to access the low bias field regime and at the same time recover the ideal of the impulsive limit without degradation of the signal and, consequently, without loss of sensitivity. Performing accurate experiments in this parameter region could be of interest for a number of sensing implementations, such as temperature, strain, or electric field, since it is at low bias field where these magnitudes have the greatest impact. Moreover, low frequency magnetic fields are better detected in such a way. Hence, our protocol represents a crucial extension of the nano-NMR scheme.

\begin{acknowledgments}
{\em Acknowledgments} ---
We thank Daniel Louzon and Alex Retzker for useful discussions. J.C. acknowledges the support from {\em Ministerio de Ciencia, Innovaci\'on y Universidades} (Spain) (``Beatriz Galindo'' Fellowship BEAGAL18/00081), S.O.C.~is supported by the \textit{Fundaci{\'o}n Ram{\'o}n Areces}  postdoctoral fellowship (XXXI edition of grants for Postgraduate Studies in Life and Matter Sciences in Foreign Universities and Research Centers 2019/2020), and J.P. is grateful for financial support from {\em Ministerio de Ciencia, Innovaci\'on y Universidades} (Spain) project PGC2018-097328-B-100, FEDER funds FIS2015-69512-R and Fundaci\'on S\'eneca (Murcia, Spain) Projects No. 19882/GERM/15. 
\end{acknowledgments}

\bibliography{pulses}

\begin{thebibliography}{50}%
\makeatletter
\providecommand \@ifxundefined [1]{%
 \@ifx{#1\undefined}
}%
\providecommand \@ifnum [1]{%
 \ifnum #1\expandafter \@firstoftwo
 \else \expandafter \@secondoftwo
 \fi
}%
\providecommand \@ifx [1]{%
 \ifx #1\expandafter \@firstoftwo
 \else \expandafter \@secondoftwo
 \fi
}%
\providecommand \natexlab [1]{#1}%
\providecommand \enquote  [1]{``#1''}%
\providecommand \bibnamefont  [1]{#1}%
\providecommand \bibfnamefont [1]{#1}%
\providecommand \citenamefont [1]{#1}%
\providecommand \href@noop [0]{\@secondoftwo}%
\providecommand \href [0]{\begingroup \@sanitize@url \@href}%
\providecommand \@href[1]{\@@startlink{#1}\@@href}%
\providecommand \@@href[1]{\endgroup#1\@@endlink}%
\providecommand \@sanitize@url [0]{\catcode `\\12\catcode `\$12\catcode
  `\&12\catcode `\#12\catcode `\^12\catcode `\_12\catcode `\%12\relax}%
\providecommand \@@startlink[1]{}%
\providecommand \@@endlink[0]{}%
\providecommand \url  [0]{\begingroup\@sanitize@url \@url }%
\providecommand \@url [1]{\endgroup\@href {#1}{\urlprefix }}%
\providecommand \urlprefix  [0]{URL }%
\providecommand \Eprint [0]{\href }%
\providecommand \doibase [0]{http://dx.doi.org/}%
\providecommand \selectlanguage [0]{\@gobble}%
\providecommand \bibinfo  [0]{\@secondoftwo}%
\providecommand \bibfield  [0]{\@secondoftwo}%
\providecommand \translation [1]{[#1]}%
\providecommand \BibitemOpen [0]{}%
\providecommand \bibitemStop [0]{}%
\providecommand \bibitemNoStop [0]{.\EOS\space}%
\providecommand \EOS [0]{\spacefactor3000\relax}%
\providecommand \BibitemShut  [1]{\csname bibitem#1\endcsname}%
\let\auto@bib@innerbib\@empty
\bibitem [{\citenamefont {Jelezko}\ \emph {et~al.}(2002)\citenamefont
  {Jelezko}, \citenamefont {Popa}, \citenamefont {Gruber}, \citenamefont
  {Tietz}, \citenamefont {Wrachtrup}, \citenamefont {Nizovtsev},\ and\
  \citenamefont {Kilin}}]{Jelezko2002}%
  \BibitemOpen
  \bibfield  {author} {\bibinfo {author} {\bibfnamefont {F.}~\bibnamefont
  {Jelezko}}, \bibinfo {author} {\bibfnamefont {I.}~\bibnamefont {Popa}},
  \bibinfo {author} {\bibfnamefont {A.}~\bibnamefont {Gruber}}, \bibinfo
  {author} {\bibfnamefont {C.}~\bibnamefont {Tietz}}, \bibinfo {author}
  {\bibfnamefont {J.}~\bibnamefont {Wrachtrup}}, \bibinfo {author}
  {\bibfnamefont {A.}~\bibnamefont {Nizovtsev}}, \ and\ \bibinfo {author}
  {\bibfnamefont {S.}~\bibnamefont {Kilin}},\ }\href {\doibase
  10.1063/1.1507838} {\bibfield  {journal} {\bibinfo  {journal} {Applied
  Physics Letters}\ }\textbf {\bibinfo {volume} {81}},\ \bibinfo {pages} {2160}
  (\bibinfo {year} {2002})}\BibitemShut {NoStop}%
\bibitem [{\citenamefont {Jelezko}\ \emph {et~al.}(2004)\citenamefont
  {Jelezko}, \citenamefont {Gaebel}, \citenamefont {Popa}, \citenamefont
  {Gruber},\ and\ \citenamefont {Wrachtrup}}]{Jelezko2004}%
  \BibitemOpen
  \bibfield  {author} {\bibinfo {author} {\bibfnamefont {F.}~\bibnamefont
  {Jelezko}}, \bibinfo {author} {\bibfnamefont {T.}~\bibnamefont {Gaebel}},
  \bibinfo {author} {\bibfnamefont {I.}~\bibnamefont {Popa}}, \bibinfo {author}
  {\bibfnamefont {A.}~\bibnamefont {Gruber}}, \ and\ \bibinfo {author}
  {\bibfnamefont {J.}~\bibnamefont {Wrachtrup}},\ }\href {\doibase
  10.1103/PhysRevLett.92.076401} {\bibfield  {journal} {\bibinfo  {journal}
  {Phys. Rev. Lett.}\ }\textbf {\bibinfo {volume} {92}},\ \bibinfo {pages}
  {076401} (\bibinfo {year} {2004})}\BibitemShut {NoStop}%
\bibitem [{\citenamefont {Doherty}\ \emph {et~al.}(2013)\citenamefont
  {Doherty}, \citenamefont {Manson}, \citenamefont {Delaney}, \citenamefont
  {Jelezko}, \citenamefont {Wrachtrup},\ and\ \citenamefont
  {Hollenberg}}]{Doherty2013}%
  \BibitemOpen
  \bibfield  {author} {\bibinfo {author} {\bibfnamefont {M.~W.}\ \bibnamefont
  {Doherty}}, \bibinfo {author} {\bibfnamefont {N.~B.}\ \bibnamefont {Manson}},
  \bibinfo {author} {\bibfnamefont {P.}~\bibnamefont {Delaney}}, \bibinfo
  {author} {\bibfnamefont {F.}~\bibnamefont {Jelezko}}, \bibinfo {author}
  {\bibfnamefont {J.}~\bibnamefont {Wrachtrup}}, \ and\ \bibinfo {author}
  {\bibfnamefont {L.~C.}\ \bibnamefont {Hollenberg}},\ }\href {\doibase
  https://doi.org/10.1016/j.physrep.2013.02.001} {\bibfield  {journal}
  {\bibinfo  {journal} {Physics Reports}\ }\textbf {\bibinfo {volume} {528}},\
  \bibinfo {pages} {1 } (\bibinfo {year} {2013})}\BibitemShut {NoStop}%
\bibitem [{\citenamefont {Childress}\ \emph {et~al.}(2006)\citenamefont
  {Childress}, \citenamefont {{Gurudev Dutt}}, \citenamefont {Taylor},
  \citenamefont {Zibrov}, \citenamefont {Jelezko}, \citenamefont {Wrachtrup},
  \citenamefont {Hemmer},\ and\ \citenamefont {Lukin}}]{Childress2006}%
  \BibitemOpen
  \bibfield  {author} {\bibinfo {author} {\bibfnamefont {L.}~\bibnamefont
  {Childress}}, \bibinfo {author} {\bibfnamefont {M.~V.}\ \bibnamefont
  {{Gurudev Dutt}}}, \bibinfo {author} {\bibfnamefont {J.~M.}\ \bibnamefont
  {Taylor}}, \bibinfo {author} {\bibfnamefont {A.~S.}\ \bibnamefont {Zibrov}},
  \bibinfo {author} {\bibfnamefont {F.}~\bibnamefont {Jelezko}}, \bibinfo
  {author} {\bibfnamefont {J.}~\bibnamefont {Wrachtrup}}, \bibinfo {author}
  {\bibfnamefont {P.~R.}\ \bibnamefont {Hemmer}}, \ and\ \bibinfo {author}
  {\bibfnamefont {M.~D.}\ \bibnamefont {Lukin}},\ }\href {\doibase
  10.1126/science.1131871} {\bibfield  {journal} {\bibinfo  {journal}
  {Science}\ }\textbf {\bibinfo {volume} {314}},\ \bibinfo {pages} {281}
  (\bibinfo {year} {2006})}\BibitemShut {NoStop}%
\bibitem [{\citenamefont {Balasubramanian}\ \emph {et~al.}(2008)\citenamefont
  {Balasubramanian}, \citenamefont {Chan}, \citenamefont {Kolesov},
  \citenamefont {Al-Hmoud}, \citenamefont {Tisler}, \citenamefont {Shin},
  \citenamefont {Kim}, \citenamefont {Wojcik}, \citenamefont {Hemmer},
  \citenamefont {Krueger}, \citenamefont {Hanke}, \citenamefont
  {Leitenstorfer}, \citenamefont {Bratschitsch}, \citenamefont {Jelezko},\ and\
  \citenamefont {Wrachtrup}}]{Balasubramanian2008}%
  \BibitemOpen
  \bibfield  {author} {\bibinfo {author} {\bibfnamefont {G.}~\bibnamefont
  {Balasubramanian}}, \bibinfo {author} {\bibfnamefont {I.~Y.}\ \bibnamefont
  {Chan}}, \bibinfo {author} {\bibfnamefont {R.}~\bibnamefont {Kolesov}},
  \bibinfo {author} {\bibfnamefont {M.}~\bibnamefont {Al-Hmoud}}, \bibinfo
  {author} {\bibfnamefont {J.}~\bibnamefont {Tisler}}, \bibinfo {author}
  {\bibfnamefont {C.}~\bibnamefont {Shin}}, \bibinfo {author} {\bibfnamefont
  {C.}~\bibnamefont {Kim}}, \bibinfo {author} {\bibfnamefont {A.}~\bibnamefont
  {Wojcik}}, \bibinfo {author} {\bibfnamefont {P.~R.}\ \bibnamefont {Hemmer}},
  \bibinfo {author} {\bibfnamefont {A.}~\bibnamefont {Krueger}}, \bibinfo
  {author} {\bibfnamefont {T.}~\bibnamefont {Hanke}}, \bibinfo {author}
  {\bibfnamefont {A.}~\bibnamefont {Leitenstorfer}}, \bibinfo {author}
  {\bibfnamefont {R.}~\bibnamefont {Bratschitsch}}, \bibinfo {author}
  {\bibfnamefont {F.}~\bibnamefont {Jelezko}}, \ and\ \bibinfo {author}
  {\bibfnamefont {J.}~\bibnamefont {Wrachtrup}},\ }\href {\doibase
  10.1038/nature07278} {\bibfield  {journal} {\bibinfo  {journal} {Nature}\
  }\textbf {\bibinfo {volume} {455}},\ \bibinfo {pages} {648} (\bibinfo {year}
  {2008})}\BibitemShut {NoStop}%
\bibitem [{\citenamefont {Maze}\ \emph {et~al.}(2008)\citenamefont {Maze},
  \citenamefont {Stanwix}, \citenamefont {Hodges}, \citenamefont {Hong},
  \citenamefont {Taylor}, \citenamefont {Cappellaro}, \citenamefont {Jiang},
  \citenamefont {Dutt}, \citenamefont {Togan}, \citenamefont {Zibrov},
  \citenamefont {Yacoby}, \citenamefont {Walsworth},\ and\ \citenamefont
  {Lukin}}]{Maze2008}%
  \BibitemOpen
  \bibfield  {author} {\bibinfo {author} {\bibfnamefont {J.~R.}\ \bibnamefont
  {Maze}}, \bibinfo {author} {\bibfnamefont {P.~L.}\ \bibnamefont {Stanwix}},
  \bibinfo {author} {\bibfnamefont {J.~S.}\ \bibnamefont {Hodges}}, \bibinfo
  {author} {\bibfnamefont {S.}~\bibnamefont {Hong}}, \bibinfo {author}
  {\bibfnamefont {J.~M.}\ \bibnamefont {Taylor}}, \bibinfo {author}
  {\bibfnamefont {P.}~\bibnamefont {Cappellaro}}, \bibinfo {author}
  {\bibfnamefont {L.}~\bibnamefont {Jiang}}, \bibinfo {author} {\bibfnamefont
  {M.~V.}\ \bibnamefont {Dutt}}, \bibinfo {author} {\bibfnamefont
  {E.}~\bibnamefont {Togan}}, \bibinfo {author} {\bibfnamefont {A.~S.}\
  \bibnamefont {Zibrov}}, \bibinfo {author} {\bibfnamefont {A.}~\bibnamefont
  {Yacoby}}, \bibinfo {author} {\bibfnamefont {R.~L.}\ \bibnamefont
  {Walsworth}}, \ and\ \bibinfo {author} {\bibfnamefont {M.~D.}\ \bibnamefont
  {Lukin}},\ }\href {\doibase 10.1038/nature07279} {\bibfield  {journal}
  {\bibinfo  {journal} {Nature}\ }\textbf {\bibinfo {volume} {455}},\ \bibinfo
  {pages} {644} (\bibinfo {year} {2008})}\BibitemShut {NoStop}%
\bibitem [{\citenamefont {Jacques}\ \emph {et~al.}(2009)\citenamefont
  {Jacques}, \citenamefont {Neumann}, \citenamefont {Beck}, \citenamefont
  {Markham}, \citenamefont {Twitchen}, \citenamefont {Meijer}, \citenamefont
  {Kaiser}, \citenamefont {Balasubramanian}, \citenamefont {Jelezko},\ and\
  \citenamefont {Wrachtrup}}]{Jacques2009}%
  \BibitemOpen
  \bibfield  {author} {\bibinfo {author} {\bibfnamefont {V.}~\bibnamefont
  {Jacques}}, \bibinfo {author} {\bibfnamefont {P.}~\bibnamefont {Neumann}},
  \bibinfo {author} {\bibfnamefont {J.}~\bibnamefont {Beck}}, \bibinfo {author}
  {\bibfnamefont {M.}~\bibnamefont {Markham}}, \bibinfo {author} {\bibfnamefont
  {D.}~\bibnamefont {Twitchen}}, \bibinfo {author} {\bibfnamefont
  {J.}~\bibnamefont {Meijer}}, \bibinfo {author} {\bibfnamefont
  {F.}~\bibnamefont {Kaiser}}, \bibinfo {author} {\bibfnamefont
  {G.}~\bibnamefont {Balasubramanian}}, \bibinfo {author} {\bibfnamefont
  {F.}~\bibnamefont {Jelezko}}, \ and\ \bibinfo {author} {\bibfnamefont
  {J.}~\bibnamefont {Wrachtrup}},\ }\href {\doibase
  10.1103/PhysRevLett.102.057403} {\bibfield  {journal} {\bibinfo  {journal}
  {Phys. Rev. Lett.}\ }\textbf {\bibinfo {volume} {102}},\ \bibinfo {pages}
  {057403} (\bibinfo {year} {2009})}\BibitemShut {NoStop}%
\bibitem [{\citenamefont {Staudacher}\ \emph {et~al.}(2013)\citenamefont
  {Staudacher}, \citenamefont {Shi}, \citenamefont {Pezzagna}, \citenamefont
  {Meijer}, \citenamefont {Du}, \citenamefont {Meriles}, \citenamefont
  {Reinhard},\ and\ \citenamefont {Wrachtrup}}]{Staudacher2013}%
  \BibitemOpen
  \bibfield  {author} {\bibinfo {author} {\bibfnamefont {T.}~\bibnamefont
  {Staudacher}}, \bibinfo {author} {\bibfnamefont {F.}~\bibnamefont {Shi}},
  \bibinfo {author} {\bibfnamefont {S.}~\bibnamefont {Pezzagna}}, \bibinfo
  {author} {\bibfnamefont {J.}~\bibnamefont {Meijer}}, \bibinfo {author}
  {\bibfnamefont {J.}~\bibnamefont {Du}}, \bibinfo {author} {\bibfnamefont
  {C.~A.}\ \bibnamefont {Meriles}}, \bibinfo {author} {\bibfnamefont
  {F.}~\bibnamefont {Reinhard}}, \ and\ \bibinfo {author} {\bibfnamefont
  {J.}~\bibnamefont {Wrachtrup}},\ }\href {\doibase 10.1126/science.1231675}
  {\bibfield  {journal} {\bibinfo  {journal} {Science}\ }\textbf {\bibinfo
  {volume} {339}},\ \bibinfo {pages} {561} (\bibinfo {year}
  {2013})}\BibitemShut {NoStop}%
\bibitem [{\citenamefont {Mamin}\ \emph {et~al.}(2013)\citenamefont {Mamin},
  \citenamefont {Kim}, \citenamefont {Sherwood}, \citenamefont {Rettner},
  \citenamefont {Ohno}, \citenamefont {Awschalom},\ and\ \citenamefont
  {Rugar}}]{Mamin2013}%
  \BibitemOpen
  \bibfield  {author} {\bibinfo {author} {\bibfnamefont {H.~J.}\ \bibnamefont
  {Mamin}}, \bibinfo {author} {\bibfnamefont {M.}~\bibnamefont {Kim}}, \bibinfo
  {author} {\bibfnamefont {M.~H.}\ \bibnamefont {Sherwood}}, \bibinfo {author}
  {\bibfnamefont {C.~T.}\ \bibnamefont {Rettner}}, \bibinfo {author}
  {\bibfnamefont {K.}~\bibnamefont {Ohno}}, \bibinfo {author} {\bibfnamefont
  {D.~D.}\ \bibnamefont {Awschalom}}, \ and\ \bibinfo {author} {\bibfnamefont
  {D.}~\bibnamefont {Rugar}},\ }\href {\doibase 10.1126/science.1231540}
  {\bibfield  {journal} {\bibinfo  {journal} {Science}\ }\textbf {\bibinfo
  {volume} {339}},\ \bibinfo {pages} {557} (\bibinfo {year}
  {2013})}\BibitemShut {NoStop}%
\bibitem [{\citenamefont {M{\"u}ller}\ \emph {et~al.}(2014)\citenamefont
  {M{\"u}ller}, \citenamefont {Kong}, \citenamefont {Cai}, \citenamefont
  {Melentijevic}, \citenamefont {Stacey}, \citenamefont {Markham},
  \citenamefont {Twitchen}, \citenamefont {Isoya}, \citenamefont {Pezzagna},
  \citenamefont {Meijer}, \citenamefont {Du}, \citenamefont {Plenio},
  \citenamefont {Naydenov}, \citenamefont {McGuinness},\ and\ \citenamefont
  {Jelezko}}]{Muller2014}%
  \BibitemOpen
  \bibfield  {author} {\bibinfo {author} {\bibfnamefont {C.}~\bibnamefont
  {M{\"u}ller}}, \bibinfo {author} {\bibfnamefont {X.}~\bibnamefont {Kong}},
  \bibinfo {author} {\bibfnamefont {J.-M.}\ \bibnamefont {Cai}}, \bibinfo
  {author} {\bibfnamefont {K.}~\bibnamefont {Melentijevic}}, \bibinfo {author}
  {\bibfnamefont {A.}~\bibnamefont {Stacey}}, \bibinfo {author} {\bibfnamefont
  {M.}~\bibnamefont {Markham}}, \bibinfo {author} {\bibfnamefont
  {D.}~\bibnamefont {Twitchen}}, \bibinfo {author} {\bibfnamefont
  {J.}~\bibnamefont {Isoya}}, \bibinfo {author} {\bibfnamefont
  {S.}~\bibnamefont {Pezzagna}}, \bibinfo {author} {\bibfnamefont
  {J.}~\bibnamefont {Meijer}}, \bibinfo {author} {\bibfnamefont {J.~F.}\
  \bibnamefont {Du}}, \bibinfo {author} {\bibfnamefont {M.~B.}\ \bibnamefont
  {Plenio}}, \bibinfo {author} {\bibfnamefont {B.}~\bibnamefont {Naydenov}},
  \bibinfo {author} {\bibfnamefont {L.~P.}\ \bibnamefont {McGuinness}}, \ and\
  \bibinfo {author} {\bibfnamefont {F.}~\bibnamefont {Jelezko}},\ }\href
  {\doibase 10.1038/ncomms5703} {\bibfield  {journal} {\bibinfo  {journal}
  {Nature Communications}\ }\textbf {\bibinfo {volume} {5}},\ \bibinfo {pages}
  {4703} (\bibinfo {year} {2014})}\BibitemShut {NoStop}%
\bibitem [{\citenamefont {Ajoy}\ \emph {et~al.}(2015)\citenamefont {Ajoy},
  \citenamefont {Bissbort}, \citenamefont {Lukin}, \citenamefont {Walsworth},\
  and\ \citenamefont {Cappellaro}}]{Ajoi2015}%
  \BibitemOpen
  \bibfield  {author} {\bibinfo {author} {\bibfnamefont {A.}~\bibnamefont
  {Ajoy}}, \bibinfo {author} {\bibfnamefont {U.}~\bibnamefont {Bissbort}},
  \bibinfo {author} {\bibfnamefont {M.~D.}\ \bibnamefont {Lukin}}, \bibinfo
  {author} {\bibfnamefont {R.~L.}\ \bibnamefont {Walsworth}}, \ and\ \bibinfo
  {author} {\bibfnamefont {P.}~\bibnamefont {Cappellaro}},\ }\href {\doibase
  10.1103/PhysRevX.5.011001} {\bibfield  {journal} {\bibinfo  {journal} {Phys.
  Rev. X}\ }\textbf {\bibinfo {volume} {5}},\ \bibinfo {pages} {011001}
  (\bibinfo {year} {2015})}\BibitemShut {NoStop}%
\bibitem [{\citenamefont {Lovchinsky}\ \emph {et~al.}(2016)\citenamefont
  {Lovchinsky}, \citenamefont {Sushkov}, \citenamefont {Urbach}, \citenamefont
  {de~Leon}, \citenamefont {Choi}, \citenamefont {De~Greve}, \citenamefont
  {Evans}, \citenamefont {Gertner}, \citenamefont {Bersin}, \citenamefont
  {M{\"u}ller}, \citenamefont {McGuinness}, \citenamefont {Jelezko},
  \citenamefont {Walsworth}, \citenamefont {Park},\ and\ \citenamefont
  {Lukin}}]{Lovchinsky2016}%
  \BibitemOpen
  \bibfield  {author} {\bibinfo {author} {\bibfnamefont {I.}~\bibnamefont
  {Lovchinsky}}, \bibinfo {author} {\bibfnamefont {A.~O.}\ \bibnamefont
  {Sushkov}}, \bibinfo {author} {\bibfnamefont {E.}~\bibnamefont {Urbach}},
  \bibinfo {author} {\bibfnamefont {N.~P.}\ \bibnamefont {de~Leon}}, \bibinfo
  {author} {\bibfnamefont {S.}~\bibnamefont {Choi}}, \bibinfo {author}
  {\bibfnamefont {K.}~\bibnamefont {De~Greve}}, \bibinfo {author}
  {\bibfnamefont {R.}~\bibnamefont {Evans}}, \bibinfo {author} {\bibfnamefont
  {R.}~\bibnamefont {Gertner}}, \bibinfo {author} {\bibfnamefont
  {E.}~\bibnamefont {Bersin}}, \bibinfo {author} {\bibfnamefont
  {C.}~\bibnamefont {M{\"u}ller}}, \bibinfo {author} {\bibfnamefont
  {L.}~\bibnamefont {McGuinness}}, \bibinfo {author} {\bibfnamefont
  {F.}~\bibnamefont {Jelezko}}, \bibinfo {author} {\bibfnamefont {R.~L.}\
  \bibnamefont {Walsworth}}, \bibinfo {author} {\bibfnamefont {H.}~\bibnamefont
  {Park}}, \ and\ \bibinfo {author} {\bibfnamefont {M.~D.}\ \bibnamefont
  {Lukin}},\ }\href {\doibase 10.1126/science.aad8022} {\bibfield  {journal}
  {\bibinfo  {journal} {Science}\ }\textbf {\bibinfo {volume} {351}},\ \bibinfo
  {pages} {836} (\bibinfo {year} {2016})}\BibitemShut {NoStop}%
\bibitem [{\citenamefont {Glenn}\ \emph {et~al.}(2018)\citenamefont {Glenn},
  \citenamefont {Bucher}, \citenamefont {Lee}, \citenamefont {Lukin},
  \citenamefont {Park},\ and\ \citenamefont {Walsworth}}]{Glenn2018}%
  \BibitemOpen
  \bibfield  {author} {\bibinfo {author} {\bibfnamefont {D.~R.}\ \bibnamefont
  {Glenn}}, \bibinfo {author} {\bibfnamefont {D.~B.}\ \bibnamefont {Bucher}},
  \bibinfo {author} {\bibfnamefont {J.}~\bibnamefont {Lee}}, \bibinfo {author}
  {\bibfnamefont {M.~D.}\ \bibnamefont {Lukin}}, \bibinfo {author}
  {\bibfnamefont {H.}~\bibnamefont {Park}}, \ and\ \bibinfo {author}
  {\bibfnamefont {R.~L.}\ \bibnamefont {Walsworth}},\ }\href {\doibase
  10.1038/nature25781} {\bibfield  {journal} {\bibinfo  {journal} {Nature}\
  }\textbf {\bibinfo {volume} {555}},\ \bibinfo {pages} {351} (\bibinfo {year}
  {2018})}\BibitemShut {NoStop}%
\bibitem [{\citenamefont {Kolkowitz}\ \emph {et~al.}(2012)\citenamefont
  {Kolkowitz}, \citenamefont {Unterreithmeier}, \citenamefont {Bennett},\ and\
  \citenamefont {Lukin}}]{Lukin2012}%
  \BibitemOpen
  \bibfield  {author} {\bibinfo {author} {\bibfnamefont {S.}~\bibnamefont
  {Kolkowitz}}, \bibinfo {author} {\bibfnamefont {Q.~P.}\ \bibnamefont
  {Unterreithmeier}}, \bibinfo {author} {\bibfnamefont {S.~D.}\ \bibnamefont
  {Bennett}}, \ and\ \bibinfo {author} {\bibfnamefont {M.~D.}\ \bibnamefont
  {Lukin}},\ }\href {\doibase 10.1103/PhysRevLett.109.137601} {\bibfield
  {journal} {\bibinfo  {journal} {Phys. Rev. Lett.}\ }\textbf {\bibinfo
  {volume} {109}},\ \bibinfo {pages} {137601} (\bibinfo {year}
  {2012})}\BibitemShut {NoStop}%
\bibitem [{\citenamefont {Taminiau}\ \emph {et~al.}(2012)\citenamefont
  {Taminiau}, \citenamefont {Wagenaar}, \citenamefont {van~der Sar},
  \citenamefont {Jelezko}, \citenamefont {Dobrovitski},\ and\ \citenamefont
  {Hanson}}]{Hanson2012}%
  \BibitemOpen
  \bibfield  {author} {\bibinfo {author} {\bibfnamefont {T.~H.}\ \bibnamefont
  {Taminiau}}, \bibinfo {author} {\bibfnamefont {J.~J.~T.}\ \bibnamefont
  {Wagenaar}}, \bibinfo {author} {\bibfnamefont {T.}~\bibnamefont {van~der
  Sar}}, \bibinfo {author} {\bibfnamefont {F.}~\bibnamefont {Jelezko}},
  \bibinfo {author} {\bibfnamefont {V.~V.}\ \bibnamefont {Dobrovitski}}, \ and\
  \bibinfo {author} {\bibfnamefont {R.}~\bibnamefont {Hanson}},\ }\href
  {\doibase 10.1103/PhysRevLett.109.137602} {\bibfield  {journal} {\bibinfo
  {journal} {Phys. Rev. Lett.}\ }\textbf {\bibinfo {volume} {109}},\ \bibinfo
  {pages} {137602} (\bibinfo {year} {2012})}\BibitemShut {NoStop}%
\bibitem [{\citenamefont {Zhao}\ \emph {et~al.}(2012)\citenamefont {Zhao},
  \citenamefont {Honert}, \citenamefont {Schmid}, \citenamefont {Klas},
  \citenamefont {Isoya}, \citenamefont {Markham}, \citenamefont {Twitchen},
  \citenamefont {Jelezko}, \citenamefont {Liu}, \citenamefont {Fedder},\ and\
  \citenamefont {Wrachtrup}}]{Zhao2012}%
  \BibitemOpen
  \bibfield  {author} {\bibinfo {author} {\bibfnamefont {N.}~\bibnamefont
  {Zhao}}, \bibinfo {author} {\bibfnamefont {J.}~\bibnamefont {Honert}},
  \bibinfo {author} {\bibfnamefont {B.}~\bibnamefont {Schmid}}, \bibinfo
  {author} {\bibfnamefont {M.}~\bibnamefont {Klas}}, \bibinfo {author}
  {\bibfnamefont {J.}~\bibnamefont {Isoya}}, \bibinfo {author} {\bibfnamefont
  {M.}~\bibnamefont {Markham}}, \bibinfo {author} {\bibfnamefont
  {D.}~\bibnamefont {Twitchen}}, \bibinfo {author} {\bibfnamefont
  {F.}~\bibnamefont {Jelezko}}, \bibinfo {author} {\bibfnamefont {R.-B.}\
  \bibnamefont {Liu}}, \bibinfo {author} {\bibfnamefont {H.}~\bibnamefont
  {Fedder}}, \ and\ \bibinfo {author} {\bibfnamefont {J.}~\bibnamefont
  {Wrachtrup}},\ }\href {\doibase 10.1038/nnano.2012.152} {\bibfield  {journal}
  {\bibinfo  {journal} {Nature Nanotechnology}\ }\textbf {\bibinfo {volume}
  {7}},\ \bibinfo {pages} {657} (\bibinfo {year} {2012})}\BibitemShut {NoStop}%
\bibitem [{\citenamefont {Laraoui}\ \emph {et~al.}(2013)\citenamefont
  {Laraoui}, \citenamefont {Dolde}, \citenamefont {Burk}, \citenamefont
  {Reinhard}, \citenamefont {Wrachtrup},\ and\ \citenamefont
  {Meriles}}]{Laraoui2013}%
  \BibitemOpen
  \bibfield  {author} {\bibinfo {author} {\bibfnamefont {A.}~\bibnamefont
  {Laraoui}}, \bibinfo {author} {\bibfnamefont {F.}~\bibnamefont {Dolde}},
  \bibinfo {author} {\bibfnamefont {C.}~\bibnamefont {Burk}}, \bibinfo {author}
  {\bibfnamefont {F.}~\bibnamefont {Reinhard}}, \bibinfo {author}
  {\bibfnamefont {J.}~\bibnamefont {Wrachtrup}}, \ and\ \bibinfo {author}
  {\bibfnamefont {C.~A.}\ \bibnamefont {Meriles}},\ }\href {\doibase
  10.1038/ncomms2685} {\bibfield  {journal} {\bibinfo  {journal} {Nature
  Communications}\ }\textbf {\bibinfo {volume} {4}},\ \bibinfo {pages} {1651}
  (\bibinfo {year} {2013})}\BibitemShut {NoStop}%
\bibitem [{\citenamefont {Dolde}\ \emph {et~al.}(2011)\citenamefont {Dolde},
  \citenamefont {Fedder}, \citenamefont {Doherty}, \citenamefont {N{\"o}bauer},
  \citenamefont {Rempp}, \citenamefont {Balasubramanian}, \citenamefont {Wolf},
  \citenamefont {Reinhard}, \citenamefont {Hollenberg}, \citenamefont
  {Jelezko},\ and\ \citenamefont {Wrachtrup}}]{Dolde2011}%
  \BibitemOpen
  \bibfield  {author} {\bibinfo {author} {\bibfnamefont {F.}~\bibnamefont
  {Dolde}}, \bibinfo {author} {\bibfnamefont {H.}~\bibnamefont {Fedder}},
  \bibinfo {author} {\bibfnamefont {M.~W.}\ \bibnamefont {Doherty}}, \bibinfo
  {author} {\bibfnamefont {T.}~\bibnamefont {N{\"o}bauer}}, \bibinfo {author}
  {\bibfnamefont {F.}~\bibnamefont {Rempp}}, \bibinfo {author} {\bibfnamefont
  {G.}~\bibnamefont {Balasubramanian}}, \bibinfo {author} {\bibfnamefont
  {T.}~\bibnamefont {Wolf}}, \bibinfo {author} {\bibfnamefont {F.}~\bibnamefont
  {Reinhard}}, \bibinfo {author} {\bibfnamefont {L.~C.~L.}\ \bibnamefont
  {Hollenberg}}, \bibinfo {author} {\bibfnamefont {F.}~\bibnamefont {Jelezko}},
  \ and\ \bibinfo {author} {\bibfnamefont {J.}~\bibnamefont {Wrachtrup}},\
  }\href {\doibase 10.1038/nphys1969} {\bibfield  {journal} {\bibinfo
  {journal} {Nature Physics}\ }\textbf {\bibinfo {volume} {7}},\ \bibinfo
  {pages} {459} (\bibinfo {year} {2011})}\BibitemShut {NoStop}%
\bibitem [{\citenamefont {Neumann}\ \emph {et~al.}(2013)\citenamefont
  {Neumann}, \citenamefont {Jakobi}, \citenamefont {Dolde}, \citenamefont
  {Burk}, \citenamefont {Reuter}, \citenamefont {Waldherr}, \citenamefont
  {Honert}, \citenamefont {Wolf}, \citenamefont {Brunner}, \citenamefont
  {Shim}, \citenamefont {Suter}, \citenamefont {Sumiya}, \citenamefont
  {Isoya},\ and\ \citenamefont {Wrachtrup}}]{Neumann2013}%
  \BibitemOpen
  \bibfield  {author} {\bibinfo {author} {\bibfnamefont {P.}~\bibnamefont
  {Neumann}}, \bibinfo {author} {\bibfnamefont {I.}~\bibnamefont {Jakobi}},
  \bibinfo {author} {\bibfnamefont {F.}~\bibnamefont {Dolde}}, \bibinfo
  {author} {\bibfnamefont {C.}~\bibnamefont {Burk}}, \bibinfo {author}
  {\bibfnamefont {R.}~\bibnamefont {Reuter}}, \bibinfo {author} {\bibfnamefont
  {G.}~\bibnamefont {Waldherr}}, \bibinfo {author} {\bibfnamefont
  {J.}~\bibnamefont {Honert}}, \bibinfo {author} {\bibfnamefont
  {T.}~\bibnamefont {Wolf}}, \bibinfo {author} {\bibfnamefont {A.}~\bibnamefont
  {Brunner}}, \bibinfo {author} {\bibfnamefont {J.~H.}\ \bibnamefont {Shim}},
  \bibinfo {author} {\bibfnamefont {D.}~\bibnamefont {Suter}}, \bibinfo
  {author} {\bibfnamefont {H.}~\bibnamefont {Sumiya}}, \bibinfo {author}
  {\bibfnamefont {J.}~\bibnamefont {Isoya}}, \ and\ \bibinfo {author}
  {\bibfnamefont {J.}~\bibnamefont {Wrachtrup}},\ }\href {\doibase
  10.1021/nl401216y} {\bibfield  {journal} {\bibinfo  {journal} {Nano Letters}\
  }\textbf {\bibinfo {volume} {13}},\ \bibinfo {pages} {2738} (\bibinfo {year}
  {2013})}\BibitemShut {NoStop}%
\bibitem [{\citenamefont {Maudsley}(1986)}]{Maudsley1986}%
  \BibitemOpen
  \bibfield  {author} {\bibinfo {author} {\bibfnamefont {A.}~\bibnamefont
  {Maudsley}},\ }\href {\doibase https://doi.org/10.1016/0022-2364(86)90160-5}
  {\bibfield  {journal} {\bibinfo  {journal} {Journal of Magnetic Resonance
  (1969)}\ }\textbf {\bibinfo {volume} {69}},\ \bibinfo {pages} {488 }
  (\bibinfo {year} {1986})}\BibitemShut {NoStop}%
\bibitem [{\citenamefont {Souza}\ \emph {et~al.}(2011)\citenamefont {Souza},
  \citenamefont {\'Alvarez},\ and\ \citenamefont {Suter}}]{Souza2011}%
  \BibitemOpen
  \bibfield  {author} {\bibinfo {author} {\bibfnamefont {A.~M.}\ \bibnamefont
  {Souza}}, \bibinfo {author} {\bibfnamefont {G.~A.}\ \bibnamefont
  {\'Alvarez}}, \ and\ \bibinfo {author} {\bibfnamefont {D.}~\bibnamefont
  {Suter}},\ }\href {\doibase 10.1103/PhysRevLett.106.240501} {\bibfield
  {journal} {\bibinfo  {journal} {Phys. Rev. Lett.}\ }\textbf {\bibinfo
  {volume} {106}},\ \bibinfo {pages} {240501} (\bibinfo {year}
  {2011})}\BibitemShut {NoStop}%
\bibitem [{\citenamefont {Wang}\ and\ \citenamefont {Liu}(2011)}]{Wang2011}%
  \BibitemOpen
  \bibfield  {author} {\bibinfo {author} {\bibfnamefont {Z.-Y.}\ \bibnamefont
  {Wang}}\ and\ \bibinfo {author} {\bibfnamefont {R.-B.}\ \bibnamefont {Liu}},\
  }\href {\doibase 10.1103/PhysRevA.83.022306} {\bibfield  {journal} {\bibinfo
  {journal} {Phys. Rev. A}\ }\textbf {\bibinfo {volume} {83}},\ \bibinfo
  {pages} {022306} (\bibinfo {year} {2011})}\BibitemShut {NoStop}%
\bibitem [{\citenamefont {Kotler}\ \emph {et~al.}(2011)\citenamefont {Kotler},
  \citenamefont {Akerman}, \citenamefont {Glickman}, \citenamefont {Keselman},\
  and\ \citenamefont {Ozeri}}]{Kotler2011}%
  \BibitemOpen
  \bibfield  {author} {\bibinfo {author} {\bibfnamefont {S.}~\bibnamefont
  {Kotler}}, \bibinfo {author} {\bibfnamefont {N.}~\bibnamefont {Akerman}},
  \bibinfo {author} {\bibfnamefont {Y.}~\bibnamefont {Glickman}}, \bibinfo
  {author} {\bibfnamefont {A.}~\bibnamefont {Keselman}}, \ and\ \bibinfo
  {author} {\bibfnamefont {R.}~\bibnamefont {Ozeri}},\ }\href {\doibase
  10.1038/nature10010} {\bibfield  {journal} {\bibinfo  {journal} {Nature}\
  }\textbf {\bibinfo {volume} {473}},\ \bibinfo {pages} {61} (\bibinfo {year}
  {2011})}\BibitemShut {NoStop}%
\bibitem [{\citenamefont {Souza}\ \emph {et~al.}(2012)\citenamefont {Souza},
  \citenamefont {Álvarez},\ and\ \citenamefont {Suter}}]{Souza2012}%
  \BibitemOpen
  \bibfield  {author} {\bibinfo {author} {\bibfnamefont {A.~M.}\ \bibnamefont
  {Souza}}, \bibinfo {author} {\bibfnamefont {G.~A.}\ \bibnamefont {Álvarez}},
  \ and\ \bibinfo {author} {\bibfnamefont {D.}~\bibnamefont {Suter}},\ }\href
  {\doibase 10.1098/rsta.2011.0355} {\bibfield  {journal} {\bibinfo  {journal}
  {Philosophical Transactions of the Royal Society A: Mathematical, Physical
  and Engineering Sciences}\ }\textbf {\bibinfo {volume} {370}},\ \bibinfo
  {pages} {4748} (\bibinfo {year} {2012})}\BibitemShut {NoStop}%
\bibitem [{\citenamefont {Cywi\ifmmode~\acute{n}\else \'{n}\fi{}ski}\ \emph
  {et~al.}(2008)\citenamefont {Cywi\ifmmode~\acute{n}\else \'{n}\fi{}ski},
  \citenamefont {Lutchyn}, \citenamefont {Nave},\ and\ \citenamefont
  {Das~Sarma}}]{Cywinski2007}%
  \BibitemOpen
  \bibfield  {author} {\bibinfo {author} {\bibfnamefont {L.}~\bibnamefont
  {Cywi\ifmmode~\acute{n}\else \'{n}\fi{}ski}}, \bibinfo {author}
  {\bibfnamefont {R.~M.}\ \bibnamefont {Lutchyn}}, \bibinfo {author}
  {\bibfnamefont {C.~P.}\ \bibnamefont {Nave}}, \ and\ \bibinfo {author}
  {\bibfnamefont {S.}~\bibnamefont {Das~Sarma}},\ }\href {\doibase
  10.1103/PhysRevB.77.174509} {\bibfield  {journal} {\bibinfo  {journal} {Phys.
  Rev. B}\ }\textbf {\bibinfo {volume} {77}},\ \bibinfo {pages} {174509}
  (\bibinfo {year} {2008})}\BibitemShut {NoStop}%
\bibitem [{\citenamefont {Uhrig}(2008)}]{Uhrig2008}%
  \BibitemOpen
  \bibfield  {author} {\bibinfo {author} {\bibfnamefont {G.~S.}\ \bibnamefont
  {Uhrig}},\ }\href {\doibase 10.1088/1367-2630/10/8/083024} {\bibfield
  {journal} {\bibinfo  {journal} {New Journal of Physics}\ }\textbf {\bibinfo
  {volume} {10}},\ \bibinfo {pages} {083024} (\bibinfo {year}
  {2008})}\BibitemShut {NoStop}%
\bibitem [{\citenamefont {Uhrig}(2007)}]{Uhrig2007a}%
  \BibitemOpen
  \bibfield  {author} {\bibinfo {author} {\bibfnamefont {G.~S.}\ \bibnamefont
  {Uhrig}},\ }\href {\doibase 10.1103/PhysRevLett.98.100504} {\bibfield
  {journal} {\bibinfo  {journal} {Phys. Rev. Lett.}\ }\textbf {\bibinfo
  {volume} {98}},\ \bibinfo {pages} {100504} (\bibinfo {year}
  {2007})}\BibitemShut {NoStop}%
\bibitem [{\citenamefont {de~Lange}\ \emph {et~al.}(2010)\citenamefont
  {de~Lange}, \citenamefont {Wang}, \citenamefont {Rist{\`e}}, \citenamefont
  {Dobrovitski},\ and\ \citenamefont {Hanson}}]{deLange2010}%
  \BibitemOpen
  \bibfield  {author} {\bibinfo {author} {\bibfnamefont {G.}~\bibnamefont
  {de~Lange}}, \bibinfo {author} {\bibfnamefont {Z.~H.}\ \bibnamefont {Wang}},
  \bibinfo {author} {\bibfnamefont {D.}~\bibnamefont {Rist{\`e}}}, \bibinfo
  {author} {\bibfnamefont {V.~V.}\ \bibnamefont {Dobrovitski}}, \ and\ \bibinfo
  {author} {\bibfnamefont {R.}~\bibnamefont {Hanson}},\ }\href {\doibase
  10.1126/science.1192739} {\bibfield  {journal} {\bibinfo  {journal}
  {Science}\ }\textbf {\bibinfo {volume} {330}},\ \bibinfo {pages} {60}
  (\bibinfo {year} {2010})}\BibitemShut {NoStop}%
\bibitem [{\citenamefont {Ryan}\ \emph {et~al.}(2010)\citenamefont {Ryan},
  \citenamefont {Hodges},\ and\ \citenamefont {Cory}}]{Ryan2010}%
  \BibitemOpen
  \bibfield  {author} {\bibinfo {author} {\bibfnamefont {C.~A.}\ \bibnamefont
  {Ryan}}, \bibinfo {author} {\bibfnamefont {J.~S.}\ \bibnamefont {Hodges}}, \
  and\ \bibinfo {author} {\bibfnamefont {D.~G.}\ \bibnamefont {Cory}},\ }\href
  {\doibase 10.1103/PhysRevLett.105.200402} {\bibfield  {journal} {\bibinfo
  {journal} {Phys. Rev. Lett.}\ }\textbf {\bibinfo {volume} {105}},\ \bibinfo
  {pages} {200402} (\bibinfo {year} {2010})}\BibitemShut {NoStop}%
\bibitem [{\citenamefont {Naydenov}\ \emph {et~al.}(2011)\citenamefont
  {Naydenov}, \citenamefont {Dolde}, \citenamefont {Hall}, \citenamefont
  {Shin}, \citenamefont {Fedder}, \citenamefont {Hollenberg}, \citenamefont
  {Jelezko},\ and\ \citenamefont {Wrachtrup}}]{Naydenov2011}%
  \BibitemOpen
  \bibfield  {author} {\bibinfo {author} {\bibfnamefont {B.}~\bibnamefont
  {Naydenov}}, \bibinfo {author} {\bibfnamefont {F.}~\bibnamefont {Dolde}},
  \bibinfo {author} {\bibfnamefont {L.~T.}\ \bibnamefont {Hall}}, \bibinfo
  {author} {\bibfnamefont {C.}~\bibnamefont {Shin}}, \bibinfo {author}
  {\bibfnamefont {H.}~\bibnamefont {Fedder}}, \bibinfo {author} {\bibfnamefont
  {L.~C.~L.}\ \bibnamefont {Hollenberg}}, \bibinfo {author} {\bibfnamefont
  {F.}~\bibnamefont {Jelezko}}, \ and\ \bibinfo {author} {\bibfnamefont
  {J.}~\bibnamefont {Wrachtrup}},\ }\href {\doibase 10.1103/PhysRevB.83.081201}
  {\bibfield  {journal} {\bibinfo  {journal} {Phys. Rev. B}\ }\textbf {\bibinfo
  {volume} {83}},\ \bibinfo {pages} {081201(R)} (\bibinfo {year}
  {2011})}\BibitemShut {NoStop}%
\bibitem [{\citenamefont {Rotem}\ \emph {et~al.}(2019)\citenamefont {Rotem},
  \citenamefont {Gefen}, \citenamefont {Oviedo-Casado}, \citenamefont {Prior},
  \citenamefont {Schmitt}, \citenamefont {Burak}, \citenamefont {McGuiness},
  \citenamefont {Jelezko},\ and\ \citenamefont {Retzker}}]{Rotem2019}%
  \BibitemOpen
  \bibfield  {author} {\bibinfo {author} {\bibfnamefont {A.}~\bibnamefont
  {Rotem}}, \bibinfo {author} {\bibfnamefont {T.}~\bibnamefont {Gefen}},
  \bibinfo {author} {\bibfnamefont {S.}~\bibnamefont {Oviedo-Casado}}, \bibinfo
  {author} {\bibfnamefont {J.}~\bibnamefont {Prior}}, \bibinfo {author}
  {\bibfnamefont {S.}~\bibnamefont {Schmitt}}, \bibinfo {author} {\bibfnamefont
  {Y.}~\bibnamefont {Burak}}, \bibinfo {author} {\bibfnamefont
  {L.}~\bibnamefont {McGuiness}}, \bibinfo {author} {\bibfnamefont
  {F.}~\bibnamefont {Jelezko}}, \ and\ \bibinfo {author} {\bibfnamefont
  {A.}~\bibnamefont {Retzker}},\ }\href {\doibase
  10.1103/PhysRevLett.122.060503} {\bibfield  {journal} {\bibinfo  {journal}
  {Phys. Rev. Lett.}\ }\textbf {\bibinfo {volume} {122}},\ \bibinfo {pages}
  {060503} (\bibinfo {year} {2019})}\BibitemShut {NoStop}%
\bibitem [{\citenamefont {Casanova}\ \emph {et~al.}(2015)\citenamefont
  {Casanova}, \citenamefont {Wang}, \citenamefont {Haase},\ and\ \citenamefont
  {Plenio}}]{Casanova2015}%
  \BibitemOpen
  \bibfield  {author} {\bibinfo {author} {\bibfnamefont {J.}~\bibnamefont
  {Casanova}}, \bibinfo {author} {\bibfnamefont {Z.-Y.}\ \bibnamefont {Wang}},
  \bibinfo {author} {\bibfnamefont {J.~F.}\ \bibnamefont {Haase}}, \ and\
  \bibinfo {author} {\bibfnamefont {M.~B.}\ \bibnamefont {Plenio}},\ }\href
  {\doibase 10.1103/PhysRevA.92.042304} {\bibfield  {journal} {\bibinfo
  {journal} {Phys. Rev. A}\ }\textbf {\bibinfo {volume} {92}},\ \bibinfo
  {pages} {042304} (\bibinfo {year} {2015})}\BibitemShut {NoStop}%
\bibitem [{\citenamefont {Abobeih}\ \emph {et~al.}(2018)\citenamefont
  {Abobeih}, \citenamefont {Cramer}, \citenamefont {Bakker}, \citenamefont
  {Kalb}, \citenamefont {Markham}, \citenamefont {Twitchen},\ and\
  \citenamefont {Taminiau}}]{Abobeih2018}%
  \BibitemOpen
  \bibfield  {author} {\bibinfo {author} {\bibfnamefont {M.~H.}\ \bibnamefont
  {Abobeih}}, \bibinfo {author} {\bibfnamefont {J.}~\bibnamefont {Cramer}},
  \bibinfo {author} {\bibfnamefont {M.~A.}\ \bibnamefont {Bakker}}, \bibinfo
  {author} {\bibfnamefont {N.}~\bibnamefont {Kalb}}, \bibinfo {author}
  {\bibfnamefont {M.}~\bibnamefont {Markham}}, \bibinfo {author} {\bibfnamefont
  {D.~J.}\ \bibnamefont {Twitchen}}, \ and\ \bibinfo {author} {\bibfnamefont
  {T.~H.}\ \bibnamefont {Taminiau}},\ }\href {\doibase
  10.1038/s41467-018-04916-z} {\bibfield  {journal} {\bibinfo  {journal}
  {Nature Communications}\ }\textbf {\bibinfo {volume} {9}},\ \bibinfo {pages}
  {2552} (\bibinfo {year} {2018})}\BibitemShut {NoStop}%
\bibitem [{\citenamefont {Wang}\ \emph {et~al.}(2019)\citenamefont {Wang},
  \citenamefont {Lang}, \citenamefont {Schmitt}, \citenamefont {Lang},
  \citenamefont {Casanova}, \citenamefont {McGuinness}, \citenamefont
  {Monteiro}, \citenamefont {Jelezko},\ and\ \citenamefont
  {Plenio}}]{Wang2019}%
  \BibitemOpen
  \bibfield  {author} {\bibinfo {author} {\bibfnamefont {Z.-Y.}\ \bibnamefont
  {Wang}}, \bibinfo {author} {\bibfnamefont {J.~E.}\ \bibnamefont {Lang}},
  \bibinfo {author} {\bibfnamefont {S.}~\bibnamefont {Schmitt}}, \bibinfo
  {author} {\bibfnamefont {J.}~\bibnamefont {Lang}}, \bibinfo {author}
  {\bibfnamefont {J.}~\bibnamefont {Casanova}}, \bibinfo {author}
  {\bibfnamefont {L.}~\bibnamefont {McGuinness}}, \bibinfo {author}
  {\bibfnamefont {T.~S.}\ \bibnamefont {Monteiro}}, \bibinfo {author}
  {\bibfnamefont {F.}~\bibnamefont {Jelezko}}, \ and\ \bibinfo {author}
  {\bibfnamefont {M.~B.}\ \bibnamefont {Plenio}},\ }\href {\doibase
  10.1103/PhysRevLett.122.200403} {\bibfield  {journal} {\bibinfo  {journal}
  {Phys. Rev. Lett.}\ }\textbf {\bibinfo {volume} {122}},\ \bibinfo {pages}
  {200403} (\bibinfo {year} {2019})}\BibitemShut {NoStop}%
\bibitem [{\citenamefont {London}\ \emph {et~al.}(2014)\citenamefont {London},
  \citenamefont {Balasubramanian}, \citenamefont {Naydenov}, \citenamefont
  {McGuinness},\ and\ \citenamefont {Jelezko}}]{London2014}%
  \BibitemOpen
  \bibfield  {author} {\bibinfo {author} {\bibfnamefont {P.}~\bibnamefont
  {London}}, \bibinfo {author} {\bibfnamefont {P.}~\bibnamefont
  {Balasubramanian}}, \bibinfo {author} {\bibfnamefont {B.}~\bibnamefont
  {Naydenov}}, \bibinfo {author} {\bibfnamefont {L.~P.}\ \bibnamefont
  {McGuinness}}, \ and\ \bibinfo {author} {\bibfnamefont {F.}~\bibnamefont
  {Jelezko}},\ }\href {\doibase 10.1103/PhysRevA.90.012302} {\bibfield
  {journal} {\bibinfo  {journal} {Phys. Rev. A}\ }\textbf {\bibinfo {volume}
  {90}},\ \bibinfo {pages} {012302} (\bibinfo {year} {2014})}\BibitemShut
  {NoStop}%
\bibitem [{\citenamefont {Reinhard}\ \emph {et~al.}(2012)\citenamefont
  {Reinhard}, \citenamefont {Shi}, \citenamefont {Zhao}, \citenamefont {Rempp},
  \citenamefont {Naydenov}, \citenamefont {Meijer}, \citenamefont {Hall},
  \citenamefont {Hollenberg}, \citenamefont {Du}, \citenamefont {Liu},\ and\
  \citenamefont {Wrachtrup}}]{Reinhard2012}%
  \BibitemOpen
  \bibfield  {author} {\bibinfo {author} {\bibfnamefont {F.}~\bibnamefont
  {Reinhard}}, \bibinfo {author} {\bibfnamefont {F.}~\bibnamefont {Shi}},
  \bibinfo {author} {\bibfnamefont {N.}~\bibnamefont {Zhao}}, \bibinfo {author}
  {\bibfnamefont {F.}~\bibnamefont {Rempp}}, \bibinfo {author} {\bibfnamefont
  {B.}~\bibnamefont {Naydenov}}, \bibinfo {author} {\bibfnamefont
  {J.}~\bibnamefont {Meijer}}, \bibinfo {author} {\bibfnamefont {L.~T.}\
  \bibnamefont {Hall}}, \bibinfo {author} {\bibfnamefont {L.}~\bibnamefont
  {Hollenberg}}, \bibinfo {author} {\bibfnamefont {J.}~\bibnamefont {Du}},
  \bibinfo {author} {\bibfnamefont {R.-B.}\ \bibnamefont {Liu}}, \ and\
  \bibinfo {author} {\bibfnamefont {J.}~\bibnamefont {Wrachtrup}},\ }\href
  {\doibase 10.1103/PhysRevLett.108.200402} {\bibfield  {journal} {\bibinfo
  {journal} {Phys. Rev. Lett.}\ }\textbf {\bibinfo {volume} {108}},\ \bibinfo
  {pages} {200402} (\bibinfo {year} {2012})}\BibitemShut {NoStop}%
\bibitem [{\citenamefont {Bauch}\ \emph {et~al.}(2018)\citenamefont {Bauch},
  \citenamefont {Hart}, \citenamefont {Schloss}, \citenamefont {Turner},
  \citenamefont {Barry}, \citenamefont {Kehayias}, \citenamefont {Singh},\ and\
  \citenamefont {Walsworth}}]{Walsworth2018}%
  \BibitemOpen
  \bibfield  {author} {\bibinfo {author} {\bibfnamefont {E.}~\bibnamefont
  {Bauch}}, \bibinfo {author} {\bibfnamefont {C.~A.}\ \bibnamefont {Hart}},
  \bibinfo {author} {\bibfnamefont {J.~M.}\ \bibnamefont {Schloss}}, \bibinfo
  {author} {\bibfnamefont {M.~J.}\ \bibnamefont {Turner}}, \bibinfo {author}
  {\bibfnamefont {J.~F.}\ \bibnamefont {Barry}}, \bibinfo {author}
  {\bibfnamefont {P.}~\bibnamefont {Kehayias}}, \bibinfo {author}
  {\bibfnamefont {S.}~\bibnamefont {Singh}}, \ and\ \bibinfo {author}
  {\bibfnamefont {R.~L.}\ \bibnamefont {Walsworth}},\ }\href {\doibase
  10.1103/PhysRevX.8.031025} {\bibfield  {journal} {\bibinfo  {journal} {Phys.
  Rev. X}\ }\textbf {\bibinfo {volume} {8}},\ \bibinfo {pages} {031025}
  (\bibinfo {year} {2018})}\BibitemShut {NoStop}%
\bibitem [{\citenamefont {Schmitt}\ \emph {et~al.}(2017)\citenamefont
  {Schmitt}, \citenamefont {Gefen}, \citenamefont {St{\"u}rner}, \citenamefont
  {Unden}, \citenamefont {Wolff}, \citenamefont {M{\"u}ller}, \citenamefont
  {Scheuer}, \citenamefont {Naydenov}, \citenamefont {Markham}, \citenamefont
  {Pezzagna}, \citenamefont {Meijer}, \citenamefont {Schwarz}, \citenamefont
  {Plenio}, \citenamefont {Retzker}, \citenamefont {McGuinness},\ and\
  \citenamefont {Jelezko}}]{Schmitt2017}%
  \BibitemOpen
  \bibfield  {author} {\bibinfo {author} {\bibfnamefont {S.}~\bibnamefont
  {Schmitt}}, \bibinfo {author} {\bibfnamefont {T.}~\bibnamefont {Gefen}},
  \bibinfo {author} {\bibfnamefont {F.~M.}\ \bibnamefont {St{\"u}rner}},
  \bibinfo {author} {\bibfnamefont {T.}~\bibnamefont {Unden}}, \bibinfo
  {author} {\bibfnamefont {G.}~\bibnamefont {Wolff}}, \bibinfo {author}
  {\bibfnamefont {C.}~\bibnamefont {M{\"u}ller}}, \bibinfo {author}
  {\bibfnamefont {J.}~\bibnamefont {Scheuer}}, \bibinfo {author} {\bibfnamefont
  {B.}~\bibnamefont {Naydenov}}, \bibinfo {author} {\bibfnamefont
  {M.}~\bibnamefont {Markham}}, \bibinfo {author} {\bibfnamefont
  {S.}~\bibnamefont {Pezzagna}}, \bibinfo {author} {\bibfnamefont
  {J.}~\bibnamefont {Meijer}}, \bibinfo {author} {\bibfnamefont
  {I.}~\bibnamefont {Schwarz}}, \bibinfo {author} {\bibfnamefont
  {M.}~\bibnamefont {Plenio}}, \bibinfo {author} {\bibfnamefont
  {A.}~\bibnamefont {Retzker}}, \bibinfo {author} {\bibfnamefont {L.~P.}\
  \bibnamefont {McGuinness}}, \ and\ \bibinfo {author} {\bibfnamefont
  {F.}~\bibnamefont {Jelezko}},\ }\href {\doibase 10.1126/science.aam5532}
  {\bibfield  {journal} {\bibinfo  {journal} {Science}\ }\textbf {\bibinfo
  {volume} {356}},\ \bibinfo {pages} {832} (\bibinfo {year}
  {2017})}\BibitemShut {NoStop}%
\bibitem [{\citenamefont {Ajoy}\ \emph {et~al.}(2019)\citenamefont {Ajoy},
  \citenamefont {Lv}, \citenamefont {Druga}, \citenamefont {Liu}, \citenamefont
  {Safvati}, \citenamefont {Morabe}, \citenamefont {Fenton}, \citenamefont
  {Nazaryan}, \citenamefont {Patel}, \citenamefont {Sjolander}, \citenamefont
  {Reimer}, \citenamefont {Sakellariou}, \citenamefont {Meriles},\ and\
  \citenamefont {Pines}}]{Ajoi2019}%
  \BibitemOpen
  \bibfield  {author} {\bibinfo {author} {\bibfnamefont {A.}~\bibnamefont
  {Ajoy}}, \bibinfo {author} {\bibfnamefont {X.}~\bibnamefont {Lv}}, \bibinfo
  {author} {\bibfnamefont {E.}~\bibnamefont {Druga}}, \bibinfo {author}
  {\bibfnamefont {K.}~\bibnamefont {Liu}}, \bibinfo {author} {\bibfnamefont
  {B.}~\bibnamefont {Safvati}}, \bibinfo {author} {\bibfnamefont
  {A.}~\bibnamefont {Morabe}}, \bibinfo {author} {\bibfnamefont
  {M.}~\bibnamefont {Fenton}}, \bibinfo {author} {\bibfnamefont
  {R.}~\bibnamefont {Nazaryan}}, \bibinfo {author} {\bibfnamefont
  {S.}~\bibnamefont {Patel}}, \bibinfo {author} {\bibfnamefont {T.~F.}\
  \bibnamefont {Sjolander}}, \bibinfo {author} {\bibfnamefont {J.~A.}\
  \bibnamefont {Reimer}}, \bibinfo {author} {\bibfnamefont {D.}~\bibnamefont
  {Sakellariou}}, \bibinfo {author} {\bibfnamefont {C.~A.}\ \bibnamefont
  {Meriles}}, \ and\ \bibinfo {author} {\bibfnamefont {A.}~\bibnamefont
  {Pines}},\ }\href {\doibase 10.1063/1.5064685} {\bibfield  {journal}
  {\bibinfo  {journal} {Review of Scientific Instruments}\ }\textbf {\bibinfo
  {volume} {90}},\ \bibinfo {pages} {013112} (\bibinfo {year}
  {2019})}\BibitemShut {NoStop}%
\bibitem [{\citenamefont {Zheng}\ \emph {et~al.}(2019)\citenamefont {Zheng},
  \citenamefont {Xu}, \citenamefont {Iwata}, \citenamefont {Lenz},
  \citenamefont {Michl}, \citenamefont {Yavkin}, \citenamefont {Nakamura},
  \citenamefont {Sumiya}, \citenamefont {Ohshima}, \citenamefont {Isoya},
  \citenamefont {Wrachtrup}, \citenamefont {Wickenbrock},\ and\ \citenamefont
  {Budker}}]{Zheng2019}%
  \BibitemOpen
  \bibfield  {author} {\bibinfo {author} {\bibfnamefont {H.}~\bibnamefont
  {Zheng}}, \bibinfo {author} {\bibfnamefont {J.}~\bibnamefont {Xu}}, \bibinfo
  {author} {\bibfnamefont {G.~Z.}\ \bibnamefont {Iwata}}, \bibinfo {author}
  {\bibfnamefont {T.}~\bibnamefont {Lenz}}, \bibinfo {author} {\bibfnamefont
  {J.}~\bibnamefont {Michl}}, \bibinfo {author} {\bibfnamefont
  {B.}~\bibnamefont {Yavkin}}, \bibinfo {author} {\bibfnamefont
  {K.}~\bibnamefont {Nakamura}}, \bibinfo {author} {\bibfnamefont
  {H.}~\bibnamefont {Sumiya}}, \bibinfo {author} {\bibfnamefont
  {T.}~\bibnamefont {Ohshima}}, \bibinfo {author} {\bibfnamefont
  {J.}~\bibnamefont {Isoya}}, \bibinfo {author} {\bibfnamefont
  {J.}~\bibnamefont {Wrachtrup}}, \bibinfo {author} {\bibfnamefont
  {A.}~\bibnamefont {Wickenbrock}}, \ and\ \bibinfo {author} {\bibfnamefont
  {D.}~\bibnamefont {Budker}},\ }\href {\doibase
  10.1103/PhysRevApplied.11.064068} {\bibfield  {journal} {\bibinfo  {journal}
  {Phys. Rev. Applied}\ }\textbf {\bibinfo {volume} {11}},\ \bibinfo {pages}
  {064068} (\bibinfo {year} {2019})}\BibitemShut {NoStop}%
\bibitem [{\citenamefont {Barry}\ \emph
  {et~al.}(2020{\natexlab{a}})\citenamefont {Barry}, \citenamefont {Schloss},
  \citenamefont {Bauch}, \citenamefont {Turner}, \citenamefont {Hart},
  \citenamefont {Pham},\ and\ \citenamefont {Walsworth}}]{Walsworth2020}%
  \BibitemOpen
  \bibfield  {author} {\bibinfo {author} {\bibfnamefont {J.~F.}\ \bibnamefont
  {Barry}}, \bibinfo {author} {\bibfnamefont {J.~M.}\ \bibnamefont {Schloss}},
  \bibinfo {author} {\bibfnamefont {E.}~\bibnamefont {Bauch}}, \bibinfo
  {author} {\bibfnamefont {M.~J.}\ \bibnamefont {Turner}}, \bibinfo {author}
  {\bibfnamefont {C.~A.}\ \bibnamefont {Hart}}, \bibinfo {author}
  {\bibfnamefont {L.~M.}\ \bibnamefont {Pham}}, \ and\ \bibinfo {author}
  {\bibfnamefont {R.~L.}\ \bibnamefont {Walsworth}},\ }\href {\doibase
  10.1103/RevModPhys.92.015004} {\bibfield  {journal} {\bibinfo  {journal}
  {Rev. Mod. Phys.}\ }\textbf {\bibinfo {volume} {92}},\ \bibinfo {pages}
  {015004} (\bibinfo {year} {2020}{\natexlab{a}})}\BibitemShut {NoStop}%
\bibitem [{\citenamefont {Cerrillo}\ \emph {et~al.}(2010)\citenamefont
  {Cerrillo}, \citenamefont {Retzker},\ and\ \citenamefont
  {Plenio}}]{Cerrillo2010}%
  \BibitemOpen
  \bibfield  {author} {\bibinfo {author} {\bibfnamefont {J.}~\bibnamefont
  {Cerrillo}}, \bibinfo {author} {\bibfnamefont {A.}~\bibnamefont {Retzker}}, \
  and\ \bibinfo {author} {\bibfnamefont {M.~B.}\ \bibnamefont {Plenio}},\
  }\href {\doibase 10.1103/PhysRevLett.104.043003} {\bibfield  {journal}
  {\bibinfo  {journal} {Phys. Rev. Lett.}\ }\textbf {\bibinfo {volume} {104}},\
  \bibinfo {pages} {043003} (\bibinfo {year} {2010})}\BibitemShut {NoStop}%
\bibitem [{\citenamefont {Cerrillo}\ \emph {et~al.}(2018)\citenamefont
  {Cerrillo}, \citenamefont {Retzker},\ and\ \citenamefont
  {Plenio}}]{Cerrillo2018}%
  \BibitemOpen
  \bibfield  {author} {\bibinfo {author} {\bibfnamefont {J.}~\bibnamefont
  {Cerrillo}}, \bibinfo {author} {\bibfnamefont {A.}~\bibnamefont {Retzker}}, \
  and\ \bibinfo {author} {\bibfnamefont {M.~B.}\ \bibnamefont {Plenio}},\
  }\href {\doibase 10.1103/PhysRevA.98.013423} {\bibfield  {journal} {\bibinfo
  {journal} {Phys. Rev. A}\ }\textbf {\bibinfo {volume} {98}},\ \bibinfo
  {pages} {013423} (\bibinfo {year} {2018})}\BibitemShut {NoStop}%
\bibitem [{\citenamefont {Fang}\ \emph {et~al.}(2013)\citenamefont {Fang},
  \citenamefont {Acosta}, \citenamefont {Santori}, \citenamefont {Huang},
  \citenamefont {Itoh}, \citenamefont {Watanabe}, \citenamefont {Shikata},\
  and\ \citenamefont {Beausoleil}}]{Watanabe2013}%
  \BibitemOpen
  \bibfield  {author} {\bibinfo {author} {\bibfnamefont {K.}~\bibnamefont
  {Fang}}, \bibinfo {author} {\bibfnamefont {V.~M.}\ \bibnamefont {Acosta}},
  \bibinfo {author} {\bibfnamefont {C.}~\bibnamefont {Santori}}, \bibinfo
  {author} {\bibfnamefont {Z.}~\bibnamefont {Huang}}, \bibinfo {author}
  {\bibfnamefont {K.~M.}\ \bibnamefont {Itoh}}, \bibinfo {author}
  {\bibfnamefont {H.}~\bibnamefont {Watanabe}}, \bibinfo {author}
  {\bibfnamefont {S.}~\bibnamefont {Shikata}}, \ and\ \bibinfo {author}
  {\bibfnamefont {R.~G.}\ \bibnamefont {Beausoleil}},\ }\href {\doibase
  10.1103/PhysRevLett.110.130802} {\bibfield  {journal} {\bibinfo  {journal}
  {Phys. Rev. Lett.}\ }\textbf {\bibinfo {volume} {110}},\ \bibinfo {pages}
  {130802} (\bibinfo {year} {2013})}\BibitemShut {NoStop}%
\bibitem [{\citenamefont {Mamin}\ \emph {et~al.}(2014)\citenamefont {Mamin},
  \citenamefont {Sherwood}, \citenamefont {Kim}, \citenamefont {Rettner},
  \citenamefont {Ohno}, \citenamefont {Awschalom},\ and\ \citenamefont
  {Rugar}}]{Mamin2014}%
  \BibitemOpen
  \bibfield  {author} {\bibinfo {author} {\bibfnamefont {H.~J.}\ \bibnamefont
  {Mamin}}, \bibinfo {author} {\bibfnamefont {M.~H.}\ \bibnamefont {Sherwood}},
  \bibinfo {author} {\bibfnamefont {M.}~\bibnamefont {Kim}}, \bibinfo {author}
  {\bibfnamefont {C.~T.}\ \bibnamefont {Rettner}}, \bibinfo {author}
  {\bibfnamefont {K.}~\bibnamefont {Ohno}}, \bibinfo {author} {\bibfnamefont
  {D.~D.}\ \bibnamefont {Awschalom}}, \ and\ \bibinfo {author} {\bibfnamefont
  {D.}~\bibnamefont {Rugar}},\ }\href {\doibase 10.1103/PhysRevLett.113.030803}
  {\bibfield  {journal} {\bibinfo  {journal} {Phys. Rev. Lett.}\ }\textbf
  {\bibinfo {volume} {113}},\ \bibinfo {pages} {030803} (\bibinfo {year}
  {2014})}\BibitemShut {NoStop}%
\bibitem [{\citenamefont {Munuera-Javaloy}\ \emph {et~al.}(2020)\citenamefont
  {Munuera-Javaloy}, \citenamefont {Arrazola}, \citenamefont {Solano},\ and\
  \citenamefont {Casanova}}]{Casanova2020}%
  \BibitemOpen
  \bibfield  {author} {\bibinfo {author} {\bibfnamefont {C.}~\bibnamefont
  {Munuera-Javaloy}}, \bibinfo {author} {\bibfnamefont {I.}~\bibnamefont
  {Arrazola}}, \bibinfo {author} {\bibfnamefont {E.}~\bibnamefont {Solano}}, \
  and\ \bibinfo {author} {\bibfnamefont {J.}~\bibnamefont {Casanova}},\ }\href
  {\doibase 10.1103/PhysRevB.101.104411} {\bibfield  {journal} {\bibinfo
  {journal} {Phys. Rev. B}\ }\textbf {\bibinfo {volume} {101}},\ \bibinfo
  {pages} {104411} (\bibinfo {year} {2020})}\BibitemShut {NoStop}%
\bibitem [{\citenamefont {Biercuk}\ \emph {et~al.}(2009)\citenamefont
  {Biercuk}, \citenamefont {Uys}, \citenamefont {VanDevender}, \citenamefont
  {Shiga}, \citenamefont {Itano},\ and\ \citenamefont
  {Bollinger}}]{Bollinger2009}%
  \BibitemOpen
  \bibfield  {author} {\bibinfo {author} {\bibfnamefont {M.~J.}\ \bibnamefont
  {Biercuk}}, \bibinfo {author} {\bibfnamefont {H.}~\bibnamefont {Uys}},
  \bibinfo {author} {\bibfnamefont {A.~P.}\ \bibnamefont {VanDevender}},
  \bibinfo {author} {\bibfnamefont {N.}~\bibnamefont {Shiga}}, \bibinfo
  {author} {\bibfnamefont {W.~M.}\ \bibnamefont {Itano}}, \ and\ \bibinfo
  {author} {\bibfnamefont {J.~J.}\ \bibnamefont {Bollinger}},\ }\href {\doibase
  10.1103/PhysRevA.79.062324} {\bibfield  {journal} {\bibinfo  {journal} {Phys.
  Rev. A}\ }\textbf {\bibinfo {volume} {79}},\ \bibinfo {pages} {062324}
  (\bibinfo {year} {2009})}\BibitemShut {NoStop}%
\bibitem [{\citenamefont {Ishikawa}\ \emph {et~al.}(2018)\citenamefont
  {Ishikawa}, \citenamefont {Yoshizawa}, \citenamefont {Mawatari},
  \citenamefont {Watanabe},\ and\ \citenamefont {Kashiwaya}}]{Ishikawa2018}%
  \BibitemOpen
  \bibfield  {author} {\bibinfo {author} {\bibfnamefont {T.}~\bibnamefont
  {Ishikawa}}, \bibinfo {author} {\bibfnamefont {A.}~\bibnamefont {Yoshizawa}},
  \bibinfo {author} {\bibfnamefont {Y.}~\bibnamefont {Mawatari}}, \bibinfo
  {author} {\bibfnamefont {H.}~\bibnamefont {Watanabe}}, \ and\ \bibinfo
  {author} {\bibfnamefont {S.}~\bibnamefont {Kashiwaya}},\ }\href {\doibase
  10.1103/PhysRevApplied.10.054059} {\bibfield  {journal} {\bibinfo  {journal}
  {Phys. Rev. Applied}\ }\textbf {\bibinfo {volume} {10}},\ \bibinfo {pages}
  {054059} (\bibinfo {year} {2018})}\BibitemShut {NoStop}%
\bibitem [{SM()}]{SM}%
  \BibitemOpen
  \href@noop {} {\bibinfo  {journal} {See suplementary material}\ }\BibitemShut
  {NoStop}%
\bibitem [{\citenamefont {Barry}\ \emph
  {et~al.}(2020{\natexlab{b}})\citenamefont {Barry}, \citenamefont {Schloss},
  \citenamefont {Bauch}, \citenamefont {Turner}, \citenamefont {Hart},
  \citenamefont {Pham},\ and\ \citenamefont {Walsworth}}]{Barry2020}%
  \BibitemOpen
\bibfield  {journal} {  }\bibfield  {author} {\bibinfo {author} {\bibfnamefont
  {J.~F.}\ \bibnamefont {Barry}}, \bibinfo {author} {\bibfnamefont {J.~M.}\
  \bibnamefont {Schloss}}, \bibinfo {author} {\bibfnamefont {E.}~\bibnamefont
  {Bauch}}, \bibinfo {author} {\bibfnamefont {M.~J.}\ \bibnamefont {Turner}},
  \bibinfo {author} {\bibfnamefont {C.~A.}\ \bibnamefont {Hart}}, \bibinfo
  {author} {\bibfnamefont {L.~M.}\ \bibnamefont {Pham}}, \ and\ \bibinfo
  {author} {\bibfnamefont {R.~L.}\ \bibnamefont {Walsworth}},\ }\href {\doibase
  10.1103/RevModPhys.92.015004} {\bibfield  {journal} {\bibinfo  {journal}
  {Rev. Mod. Phys.}\ }\textbf {\bibinfo {volume} {92}},\ \bibinfo {pages}
  {015004} (\bibinfo {year} {2020}{\natexlab{b}})}\BibitemShut {NoStop}%
\end{thebibliography}%

\newpage
\newpage

\onecolumngrid

\renewcommand\thefigure{S\arabic{figure}}
\renewcommand\theequation{S\arabic{equation}}

\section{Supplemental Information}

\section{Analytics for the Filter Function \label{app}}
\setcounter{figure}{0} 

The influence that any pulse sequence has on the system can be described through the filter function, defined as $F(\omega) = |f(\omega)|^2=f^2_r(\omega)+f^2_i(\omega)$, where $f(\omega)=f_r(\omega)+i f_i(\omega)$ is the Fourier transform $\mathcal{F}[h(t)]$ of the response function $h(t)$ of the system to a sequence of pulses. We approximate the response function with a piecewise-defined function that aims to reproduce the value of coherence at each moment of time $t$. In the interval time between pulses, $h(t)$ takes alternatively a value of $+ 1$ or $-1$. The effect of the pulses themselves is described through linear ramps connecting these values. Pulses have a temporal duration of $T$ for the $\frac{\pi}{2}$ pulses that begin and end the dynamical decoupling sequence, and 2T for the $\pi$ pulses in between (see Fig.2 in the main manuscript for an example).

The Ramsey sequence is the simplest DD protocol, which comprises only two $\frac{\pi}{2}$ pulses. We approximate its response function by
\begin{equation}
h_R(t)\propto 
\begin{cases}
0,  &t \leqslant 0 \\
\frac{t}{T}, &0 < t \leqslant T \\
1, & T < t \leqslant \tau - T \\
\frac{\tau-t}{T}, & \tau - T < t \leqslant \tau \\
0, & t > \tau.
\end{cases}\label{respfunc2}
\end{equation}
The associated filter function reads
\be
F_R(\omega,T) = \frac{16 }{T^2 \w^4} \sin ^2\left(\w\frac{T}{2}\right) \sin ^2\left(\w\frac{\tau-T}{2}\right)= \left(\tau-T\right)^2  \sinc^2\left(\w\frac{T}{2}\right)\sinc^2\left(\w\frac{\tau-T}{2}\right),
\ee
where the second expression makes use of $\sinc x = \frac{\sin x}{x}$. The expression in the impulsive limit (Eq.5 in the main manuscript) can be obtained by setting $T\rightarrow 0$. 

\begin{figure}[h!]
    \includegraphics[width=5cm]{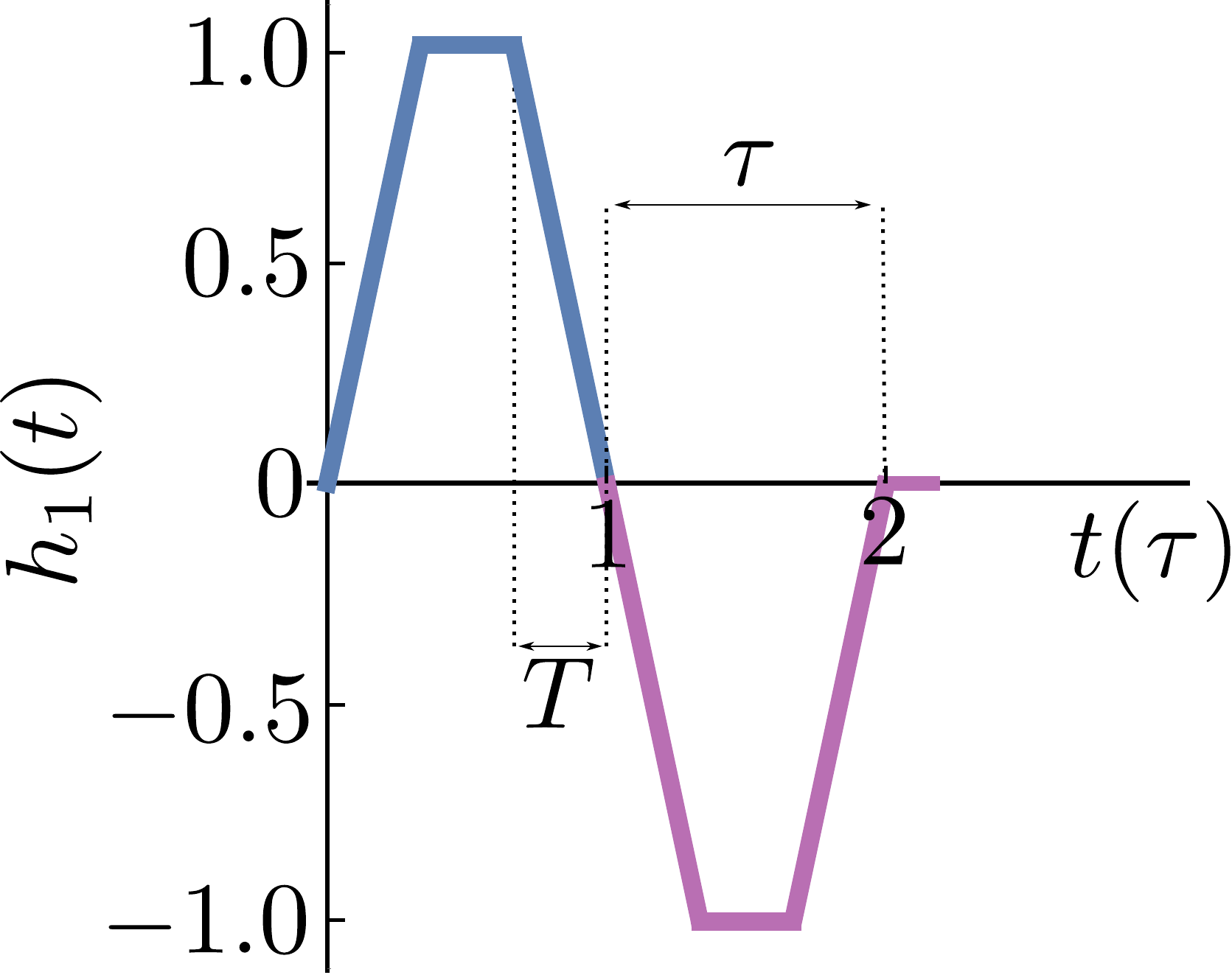}
    \caption{Schematic depiction of the response function $h_1(t)$ (Hahn echo sequence). Each segmented colour represents the response function of the Ramsey scheme $h_R(t)$. The total length of the dynamical decoupling sequence is $2\tau$ while the length of a $\frac{\pi}{2}$ pulse (or half a $\pi$ pulse) is $T$, which may take values between $0$ and $\tau/2$.}
    \label{hahnpulse}
\end{figure}

More complex DD sequences consist in the introduction of $N$ equally spaced $\pi$ pulses. The simplest case $N=1$ is known as the Hahn echo. The corresponding response function $h_1(t)$ can be interpreted as the effect of two consecutive Ramsey sequences Eq. \ref{respfunc2} of opposite sign, as shown in Fig. \ref{hahnpulse}. Thus, $h_1(t) = h_R(t) - h_R(t-\tau)$. The associated filter function can be easily derived from the Ramsey sequence filter function: $F_1(\omega,T) = |1 - \exp{i\omega\tau}|^2F_R(\omega,T)$, where we have used the time-shift property of the Fourier transform $\mathcal{F}[h(t-t')] = \exp{(i\omega t')}\mathcal{F}[h(t)]$. The filter function for the Hahn echo sequence finally reads
\be
F_1(\omega,T) = 4\sin^2\left(\w\frac{\tau}{2}\right)F_R(\omega,T).
\ee

\begin{figure}[h]
    \includegraphics[width=14cm]{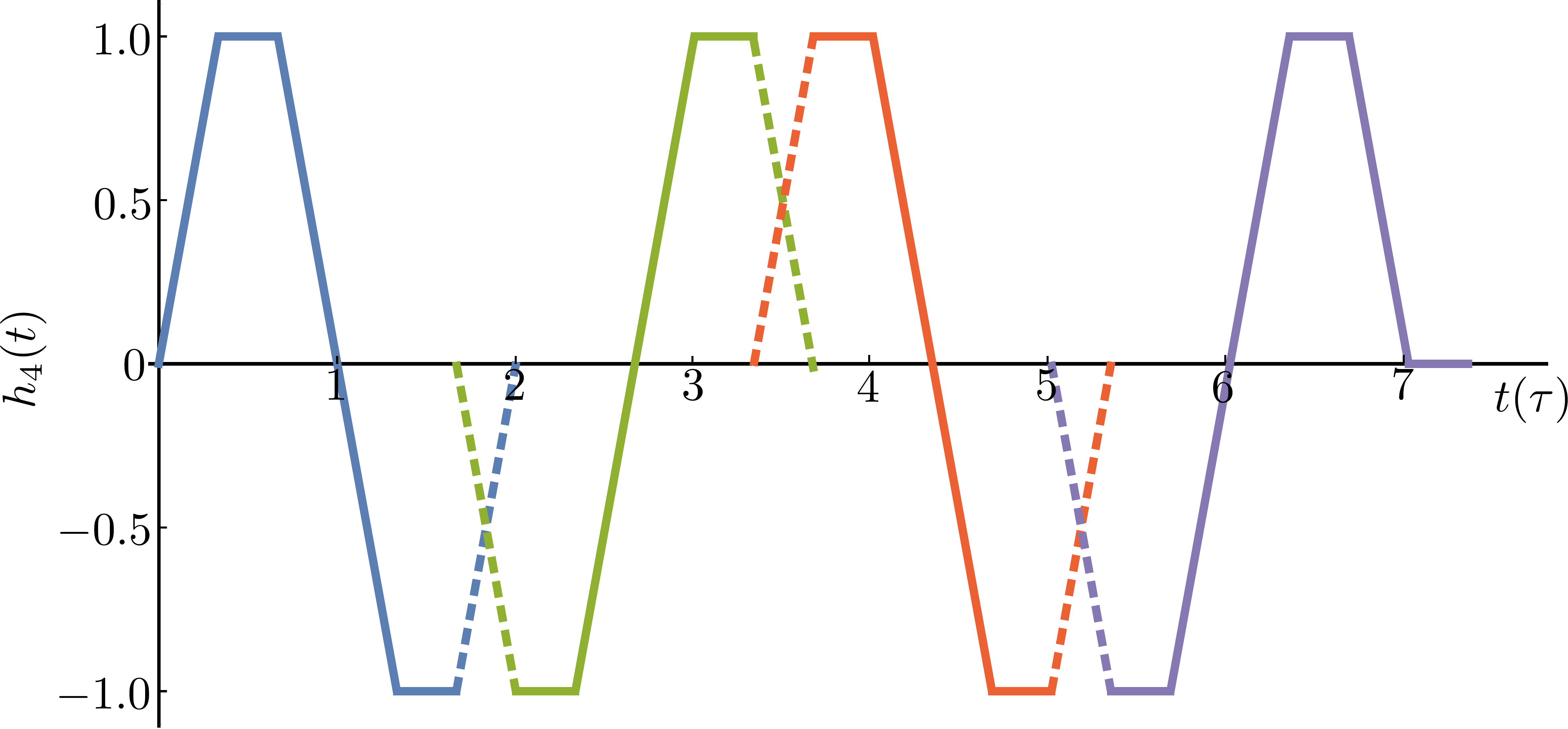}
    \caption{Schematic depiction of the response function $h_N(t)$ for $N = 4$ pulses. Each segmented colour represents the response function of the Hahn echo scheme $h_1(t)$. The total length of the dynamical decoupling sequence is $2N\tau - (N-1)T$ while the length of a $\pi/2$ pulse (or half a $\pi$ pulse) is $T$.}
    \label{pulseramps}
\end{figure}

Analogously, DD sequences with a higher number of $\pi$ pulses can be interpreted as repetitions of Hahn echo sequences of alternating sign, as we show in Fig. \ref{pulseramps}. This time we introduce an overlap of length $T$ between Hahn-echo sequences to compensate for the loss of signal as a function of $T$ of the Ramsey sequence. Thus, $h_N(t) = \sum_{j=0}^{N-1} (-1)^{j}h_1\left[t - j(2\tau-T)\right]$. The corresponding filter function can now be calculated from the Hahn echo sequence result making use of the time-shift property: $F_N(\omega,T) = |\sum_{j=0}^{N-1} (-1)^{j} \exp{(i\omega j(2\tau-T))}|^2F_1(\omega,T)$. We further make use of the sum of the geometric series $\sum_{j=0}^{N-1} x^j = \frac{1-x^{N}}{1-x}$ to finally obtain
\begin{equation}
 F_{2n+1}(\omega,T) =4\sin^2\left(\omega \frac{\tau}{2}\right)\frac{\cos^2\left(N\omega\frac{2\tau-T}{2}\right)}{\cos^2\left(\omega\frac{2\tau-T}{2}\right)} F_{R}(\omega,T),
\end{equation}
for an odd amount $N=2n+1$ of $\pi$-pulses and
\begin{equation}
 F_{2n}(\omega,T) =4\sin^2\left(\omega \frac{\tau}{2}\right)\frac{\sin^2\left(N\omega\frac{2\tau-T}{2}\right)}{\cos^2\left(\omega\frac{2\tau-T}{2}\right)} F_{R}(\omega,T),
\end{equation}
for an even amount $N=2n>0$ of $\pi$-pulses.

The accumulated phase in the NV can be calculated as $\phi(t) = \mu \intop_{-\tau}^{\tau} \text{d} a \, h(a) B (t+a)$. It is immediate then to see that for a coherent signal of the form $\mathbf{B}_r'(\w)=\cos(\w t)(0,0,1)$ [$\mathbf{B}_i'(\w)=\sin(\w t)(0,0,1)$], the accumulated phase corresponds with the cosine [sine] Fourier transform of the response function. The sum of absolute values squared then is equivalent to $F_N(\omega,T)$ calculated analytically above and consequently permits comparing directly the numerical results for accumulated phase with filter functions, where both quantify in a fundamental way the performance of the corresponding control protocols.

\section{Double Quantum Method}

\begin{figure}[h]
    \includegraphics[width=7cm]{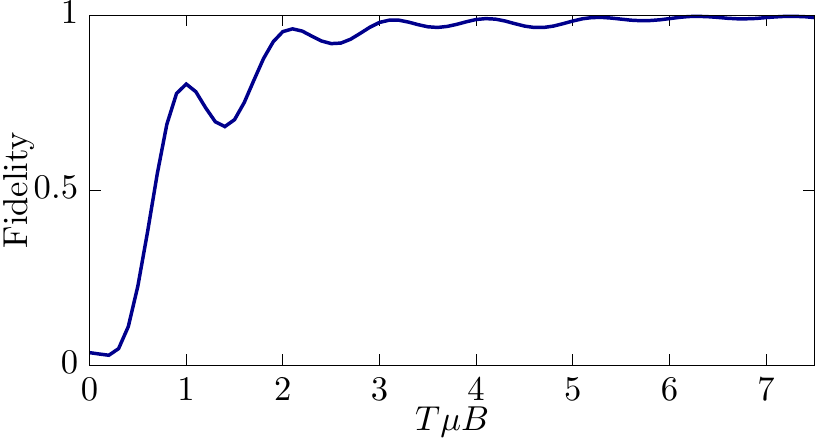}
    \caption{Fidelity of state $\left(\ket{-1} + i\ket{+1}\right)/\sqrt{2}$  represented as a function of the dimensionless quantity $T\mu B$ after the double-quantum preparation protocol. In consonance with results for the CC case, for small Zeeman splittings or short pulse lengths, poor fidelities are achieved due to off-resonant excitation of the unwanted states.}
    \label{FidDQ}
\end{figure}

In this section we show that the conclusions regarding the failure of the CC protocol in the $\mu B T< 1$ regime also apply to a more sophisticated control strategy as is the double quantum (DQ) control protocol. DQ is analogous to CC but for the fact that a superposition of states $\ket{-1}$ and $\ket{+1}$ is used as opposed to $\ket{-1}$ and $\ket{0}$. A simple strategy to achieve this state consists in the subsequent application of two pulses in resonance with the $\ket{0}\leftrightarrow\ket{-1}$ and the the $\ket{0}\leftrightarrow\ket{+1}$ transitions respectively: a microwave $\pi/2$ pulse of duration $T$ and frequency $\nu=D-\mu B$ followed by a $\pi$ pulse of duration $2T$ of frequency $\nu=D+\mu B$. Just as with CC, the pulse sequence is applied on an NV-center initialized in state $\ket{0}$. In line with results for CC shown in Fig.1b in the main text, the preparation strategy fails to reach satisfactory fidelity values for $\mu B T< 1$. Furthermore, the accumulation of error produces even worse values than in the CC case.

\section{Numerical Hahn-Echo Filter Functions}

In this section we numerically analyze the effect of the pulse length $T$ in the context of a Hahn-echo experiment in the language of FF. This data is then used to produce Fig.5 in the main text.

\begin{figure}[h]
    \includegraphics{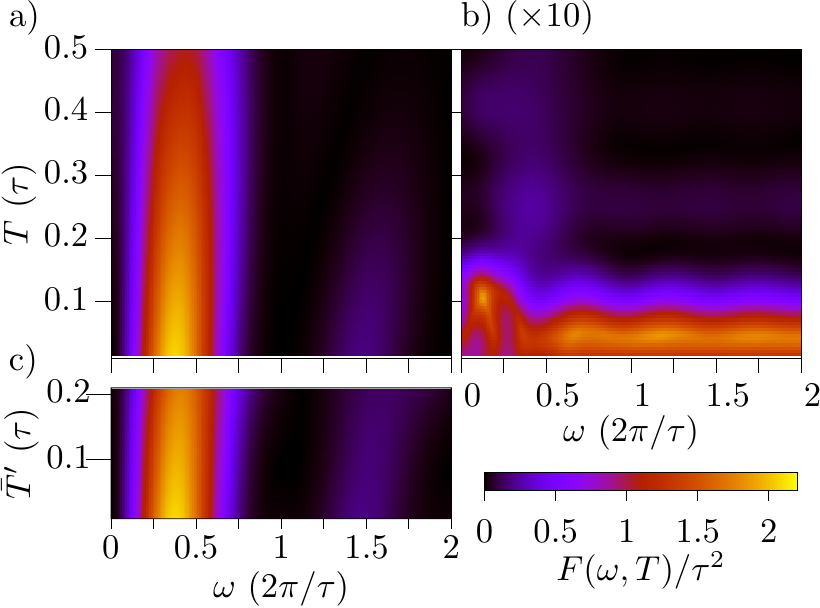}
    \caption{Filter functions $F(\w,T)$ of a Hahn echo experiment as a function of the pulse duration and the signal frequency $\w$ a) in the ideal case of a 2LS, b) for an NV center (Eq.1 in the main text) using CC and c) using NV-ERC. The value of the Zeeman splitting is $\mu B= 10/\tau$, which imposes $\bar{T}'\leq \tau \pi /10\sqrt{2}\simeq0.22\tau$. The values in b) are scaled and correspond to the function $F(\w,T)/10\tau^2$.  In c), the $\pi$ pulse has a duration $T'=0.01\tau$.}
    \label{HahnEcho}
\end{figure}

In addition to the effects discussed in the main text for a Ramsey experiment, the pulse length $T$ disturbs the value of the frequency that the Hahn-echo experiment addresses. This behavior is reproduced by the numerical calculation in an ideal 2LS as shown in Fig.\ref{HahnEcho}a. Additionally, as in Fig.4 of the main text, the same sequence on an NV center modeled by Eq.1 fails in the region $T\mu B \leq 1$ (see Fig.\ref{HahnEcho}b) to the extent that the signal is off-scale by an order of magnitude. With NV-ERC, the Hahn-echo profile is successfully recovered in that regime, as shown in Fig.\ref{HahnEcho}c. As expected from Eq.7 in the main text for $N=1$, and just like Fig.\ref{HahnEcho}a, it reproduces minima in the vicinity of $\w=n2\pi/\tau$, with $n=0,1,2\dots$, and maxima around the solutions of equation $\w\tau=\tan(\w\tau/2)$, $\w\simeq0.37\times 2\pi/\tau$ and $\w\simeq1.47\times 2\pi/\tau$.

\section{Coherence time $T_2$, sensitivity and filter functions}

We may use the derivations in \cite{Cywinski2007} to relate the concept of filter function $F\left(\w\right)$ to the coherence time $T_2$ of the NV center under a certain DD sequence. Following the derivations therein, we may express the coherence $W(t)$ of two states $\ket{a}$ and $\ket{b}$ of the NV center in terms of its negative logarithm $\chi(t)$
\be
W(t)=\frac{\left|\bra{a}\rho(t)\ket{b}\right|}{\left|\bra{a}\rho(0)\ket{b}\right|}=e^{-\chi(t)}.
\ee
In the Gaussian approximation, the function $\chi(t)$ may be expressed in terms of the filter function $F\left(\w,t\right)$ of an experiment of duration $t$ and the spectral density of the environmental noise $S(\w)$ as
\be
\chi(t)=\int_0^\infty \frac{d\w}{2\pi} S(\w) F(\w,t).
\ee
In terms of a linear approximation for $\chi(t) \simeq t/T_2$ we then obtain the relationship
\be
T_2\simeq\frac{t}{\chi(t)},
\label{eq:T2}
\ee
which is the relationship that has been used for the estimations in the context of a class of environmental noises in Fig.5 of the main text.

For the case of a flat spectral density $S(\w)=2 \pi S_0$, the integral over frequency of the filter function is related to the function $W(t)$ through
\be
W(t)=e^{-S_0\int_0^\infty  F(\w,t) d\w}
\ee
In the case of an ideal pulse sequence, this approximation always yields $T_2=2/S_0$. Nevertheless, numerically computed filter functions do not share this property, as shown in the main text.

Therefore, Eq.\ref{eq:T2} directly relates numerically computed FF (such as those shown in Fig.4 and Fig.5 of the main text) to $T_2$ times of both CC and NV-ERC for a flat spectral density. Fig.4-left and Fig.5b), corresponding to the numerical FF of CC, predicts large signal across the whole spectrum as the pulse length is reduced. Accordingly, through Eq.(\ref{eq:T2}), $T_2$ times abruptly diminish in this regime. This effect is not expected with NV-ERC in the light of Fig.4-right and Fig.5c), i.e., $T_2$ times will remain similar for the range of pulse duration.

Sensitivity $\eta$ is closely related to coherence times, as reviewed in \cite{Barry2020}. Specifically, we may use the relationship
\be
\eta=\frac{\eta_{opt}}{e^{-\tau/T_2}},
\ee
to estimate the effect that a reduced coherence time may have in the optimal, fully coherent sensitivity $\eta_{opt}$.

\section{Robustness to pulse errors}

Here we analyze the effect that imperfections on the pulses have on the NV-ERC control, and compare them with the effect that such imperfections have on conventional control sequences. We show that, as opposed to CC, NV-ERC is robust to phase, whereas duration or amplitude errors can be easily detected.

{\em Phase errors} --- Here we show that microwave pulses that address the zero-field transition of NV-centers are protected against initial phase inaccuracies. Let us consider a microwave pulse of the form
\be
H_{p}=\Omega\cos\left(Dt+\phi\right)S_{x}
\ee
Transformation into the interaction picture with respect to $DS_{z}^{2}$ yields
\be
\tilde{H}_{p}=\frac{\Omega}{\sqrt{2}}\cos\left(Dt+\phi\right)e^{iDt}\left(\left|1\right\rangle \left\langle 0\right|+\left|-1\right\rangle \left\langle 0\right|\right)+H.c.
\ee
and dropping time-dependent terms
\be
\tilde{H}_{p}=\frac{\Omega}{2\sqrt{2}}e^{-i\phi}\left(\left|1\right\rangle \left\langle 0\right|+\left|-1\right\rangle \left\langle 0\right|\right)+H.c.
\ee
or, in terms of state $\left|+\right\rangle$
\be
 \tilde{H}_{p}=e^{-i\phi}\frac{\Omega}{2}\left|+\right\rangle \left\langle 0\right|+H.c..
\ee
The phase can be reabsorbed to state $\left|0\right\rangle$  as long as the transformation either begins or ends in that state, which is always the case in NV-ERC.

\begin{figure}[h]
    \includegraphics[width=7cm]{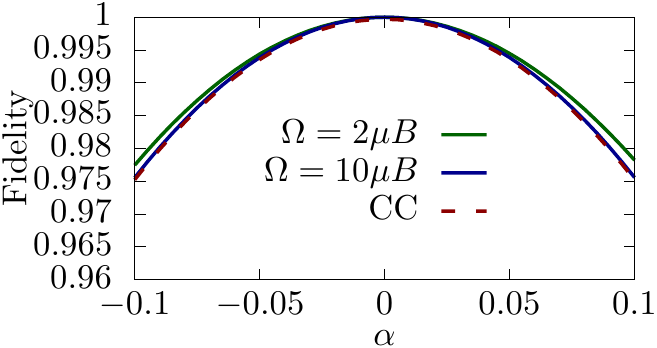}
    \caption{Fidelity of the target state for a realization of the corresponding pulse with an incorrect Rabi frequency $\W(1+\alpha)$ as a function of the unitless error $\alpha$. We consider the CC case and two NV-ERC cases of different Rabi frequency $\W$.}
    \label{RabiErr}
\end{figure}

{\em Duration and Amplitude errors} --- The preparation pulses of NV-ERC is at most as susceptible to duration and amplitude errors as CC $\pi$ pulses, as illustrated in Fig.\ref{RabiErr}.
As is the case with CC, errors can be mitigated by the use of pulse shaping and pulse correcting techniques, or with composite pulses sequences that self-correct possible individual pulses errors.

Nevertheless, they incorporate a simple test that is unavailable for CC and that certifies that the generated state is located in the equator of the $\ket{+1}-\ket{-1}$ Bloch sphere. A measurement of vanishing population in state $\ket{0}$ after the pulse is indeed identically associated to the generation of a state of the form $\ket{\phi}$. This can be exploited in experimental implementations to accurately adjust amplitude and duration of the pulses. No similarly simple test exists in the case of conventional pulses.

\end{document}